\documentclass[prl,amssymb,amsmath,superscriptaddress,twocolumn]{revtex4-2}
\usepackage[english]{babel}
\usepackage[utf8]{inputenc}
\usepackage{graphicx}
\graphicspath{{./images/}}
\usepackage{url}
\usepackage{hyperref}
\usepackage{color}
\usepackage{xcolor}
\usepackage{cancel}
\usepackage{enumerate}
\usepackage{enumitem}
\usepackage{pifont}
\usepackage{mathtools}
\usepackage{bbold}
\usepackage{lipsum}
\usepackage{tikz}
\usepackage{tikz-3dplot}
\usetikzlibrary{positioning}
\usetikzlibrary{backgrounds}
\usetikzlibrary{arrows.meta}
\usepackage{bm, bbm}
\usepackage[normalem]{ulem}
\usepackage{subcaption}
\usepackage{multirow}

\usepackage[defaultcolor=blue]{changes}
\definechangesauthor[name=Dibya, color=cyan]{d}
\definechangesauthor[name=Adam, color=purple]{a}
\definechangesauthor[name=Herb, color=orange]{h}

\newcommand{\subfigimg}[5][,]{%
  \setbox1=\hbox{\includegraphics[#1]{#3}}
  \leavevmode\rlap{\usebox1}
  \rlap{\hspace*{#4}\raisebox{\dimexpr\ht1-#5\baselineskip}{#2}}
  \phantom{\usebox1}
}

\newcommand{\dash}{%
  \hspace{-1em}\textemdash
}
\begin{document}
\title{Impact of a Lifshitz Transition on the onset of spontaneous coherence}

\author{Adam Eaton}
\affiliation{Department of Physics, Indiana University, Bloomington, Indiana 47405, USA}

\author{Dibya Mukherjee}
\affiliation{Department of Physics, Indiana University, Bloomington, Indiana 47405, USA}

\author{H. A. Fertig}
\affiliation{Department of Physics, Indiana University, Bloomington, Indiana 47405, USA}
\affiliation{Quantum Science and Engineering Center, Indiana University, Bloomington, Indiana 47405, USA}

\date{\today} 

\begin{abstract}
Lifshitz transitions are topological transitions of a Fermi surface, whose signatures typically appear in the conduction properties of a host metal. Here, we demonstrate, using an extended Falicov-Kimball model of a two-flavor fermion system, that a Lifshitz transition which occurs in the non-interacting limit impacts interaction-induced insulating phases, even though they do not host Fermi surfaces.
For strong interactions we find a first order transition between states of different polarization
This transition line ends in a very unusual quantum critical endpoint, whose presence is stabilized by the onset of inter-flavor coherence.  We demonstrate that the surfaces of maximum coherence in these states reflect the distinct Fermi surface topologies of the states separated by the non-interacting Lifshitz transition.
The phenomenon is shown to be independent of the band topologies involved. Experimental realizations of our results are discussed for both electronic and optical lattice systems.
\end{abstract}


\maketitle
\paragraph*{Introduction.}\dash
In recent years, topology has become increasingly appreciated in condensed matter physics as a framework for understading diverse physical phenomena.  These include the Thouless pump \cite{Kraus2011, Madsden2013}, topological defects \cite{Ringel2012, Teo2013, Veldhorst2012, Fu2008, Fu2009, Nelson2002, Chaikin1995}, quantized Hall effects \cite{TKKN_1982,Vayrynen2013, Hart2014, Weeks2011, Hou2009, Du2015, Bozkurt2018, Bernevig_2006, Girvin_2019}, magnetic breakdown \cite{Lu2014, Chapai2023, Alexandradinata2017, Lemut2020},  and topological insulators \cite{Hassan_2010,Qi_2011,Cooper2012, Wang2015, Neupert2011, Rachel2016, Ren2020, Lunde2013, Seshadri2019, Liu2010, Asboth2016}. In addition to its utility for theoretical understanding, topology is physically significant because it leads to phenomena that are robust with respect to various perturbations \cite{RMP2008, Lohse2018, Hu2021, Abanov2017, Hamma2013}.

Lifshitz transitions \cite{Volovik2018, Li2022, Chen2012, Lemonik2010, Akzyanov2021, Balents2016} are an example of this.  They occur when the topology of a Fermi surface changes with system parameters such as pressure, doping or external magnetic field \cite{Li2022, Chen2012, Lemonik2010, Balents2016}, and typically are observable as anomalies in  magneto-oscillation periods as the system passes through such transitions. Because a Lifshitz transition is a Fermi surface phenomenon, its impact is normally only expected in metallic systems.  In this work, we demonstrate that such transitions can also impact systems outside their metallic regime.  In particular,
we report on a system where a Lifshitz transition in the non-interacting limit leaves a clear signature when gap-opening interactions eliminate the Fermi surface.

\begin{figure}
  \centering
\begin{subfigure}{0.285\linewidth}
     \subfigimg[width=\linewidth]{\textcolor{white}{(a)}}{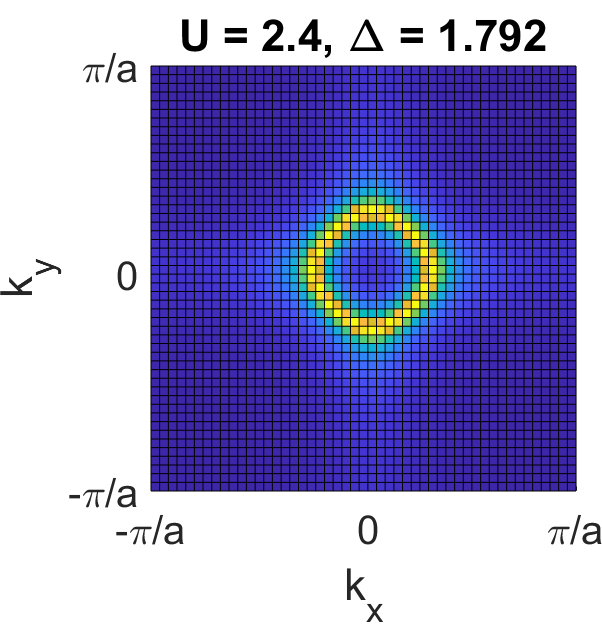}{0.7cm}{1.5}
     \refstepcounter{subfigure}\label{fig:CohI}
  \end{subfigure}
  \begin{subfigure}{0.285\linewidth}
    \subfigimg[width=\linewidth]{\textcolor{white}{(b)}}{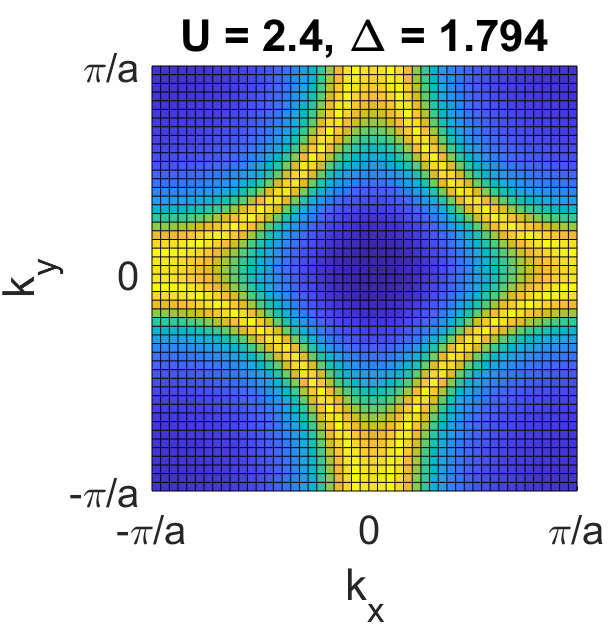}{0.7cm}{1.5}
    \refstepcounter{subfigure}\label{fig:CohII}
  \end{subfigure}
    \begin{subfigure}{0.395\linewidth}
    \subfigimg[width=\linewidth]{\textcolor{white}{(c)}}{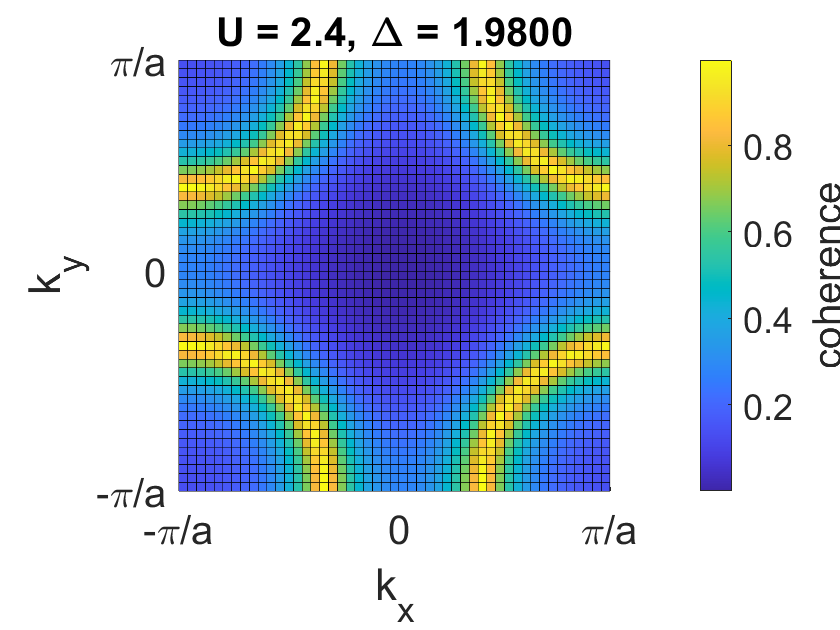}{0.8cm}{1.5}
    \refstepcounter{subfigure}\label{fig:CohIIcorner}
  \end{subfigure}
\hfill
\begin{subfigure}{0.41\linewidth}
    \subfigimg[width=\linewidth]{\textbf{(d)}}{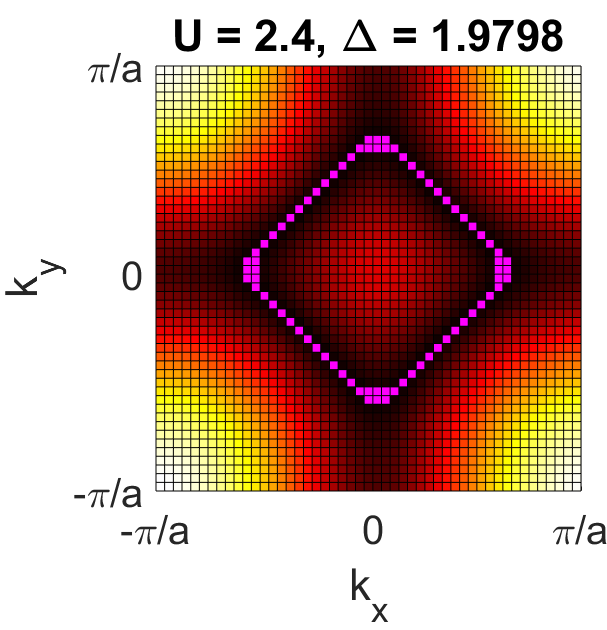}{1.0cm}{1.8}
    \refstepcounter{subfigure}\label{fig:CohIFS}
  \end{subfigure}
\begin{subfigure}{0.57\linewidth}
    \subfigimg[width=\linewidth]{\textcolor{white}{\textbf{(e)}}}{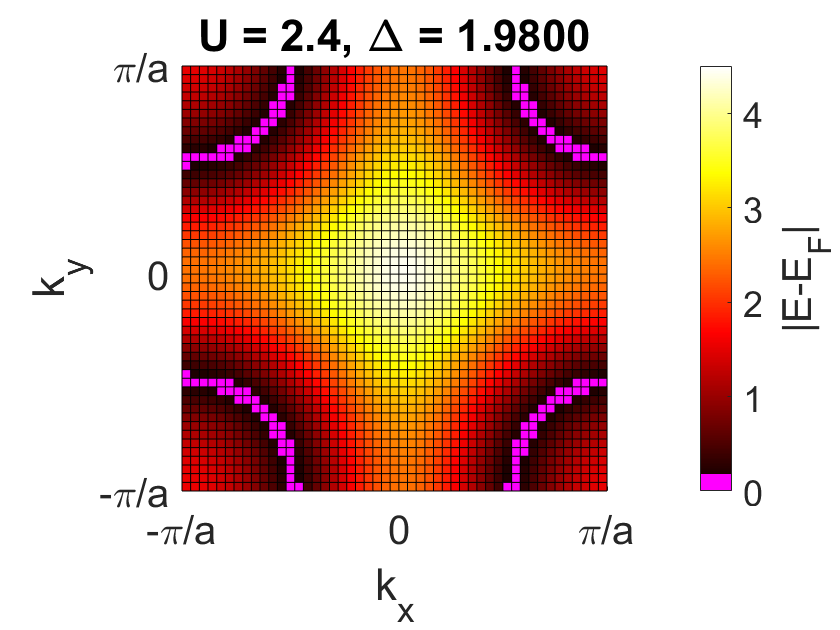}{1.2cm}{1.8}
    \refstepcounter{subfigure}\label{fig:CohIIFS}
  \end{subfigure}
        \caption{(\subref{fig:CohI}), (\subref{fig:CohII}), (\subref{fig:CohIIcorner}): the coherence as a function of crystal momentum. Regions of largest coherence form topologically distinct curves. (\subref{fig:CohIFS}) and (\subref{fig:CohIIFS}): Fermi surfaces on either side of the Lifshitz transition in absence of spontaneous coherence. }
        \label{fig:Lifshitz}
\end{figure}

\begin{figure*}[hbtp]
  \centering
\begin{subfigure}{0.49\linewidth}
    \centering
\begin{subfigure}{0.48\linewidth}
    \centering
\subfigimg[width=\textwidth]{(a)}{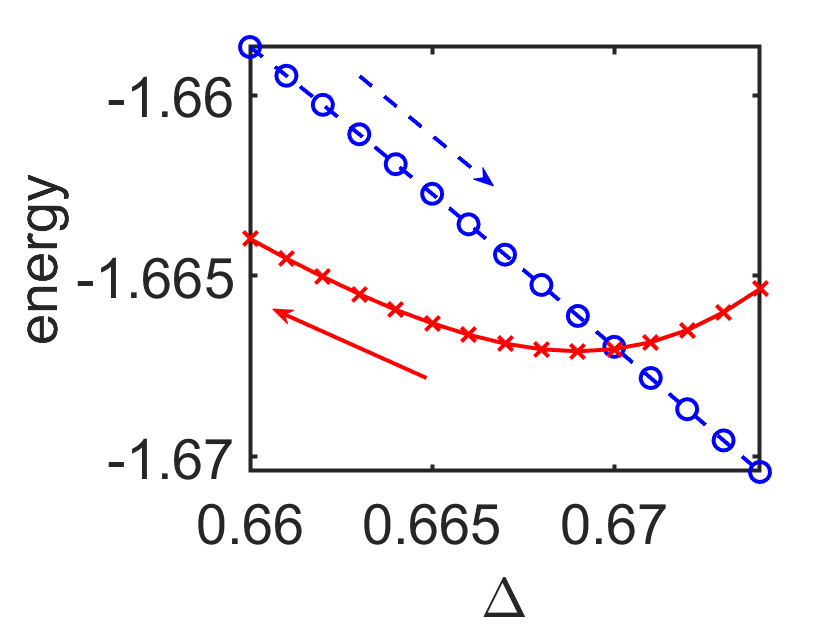}{3.2cm}{1.7}
\refstepcounter{subfigure}\label{fig:U2pt2energy}
\end{subfigure}
\begin{subfigure}{0.48\linewidth}
    \centering
\subfigimg[width=\textwidth]{(b)}{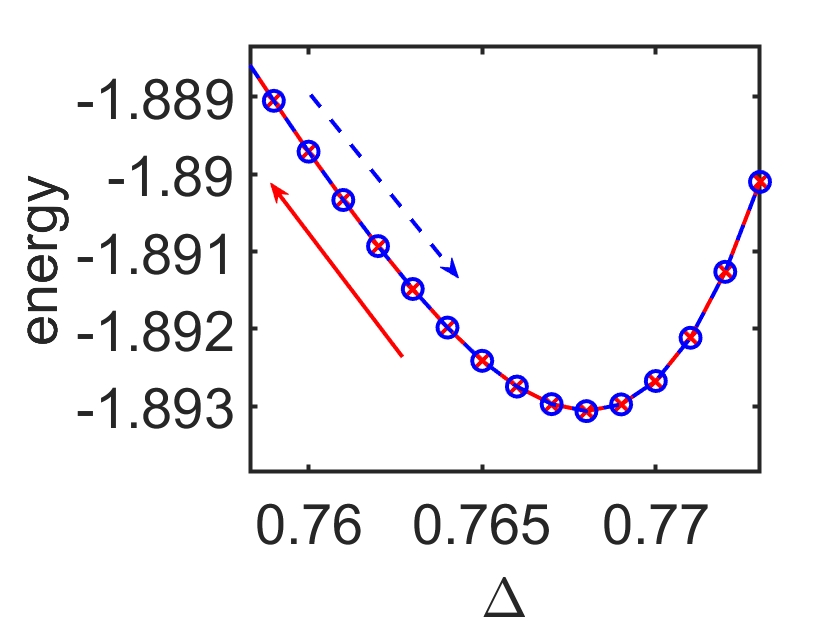}{3.2cm}{1.7}
\refstepcounter{subfigure}\label{fig:U2pt0energy}
  \end{subfigure}

\begin{subfigure}{\linewidth}
    \centering
\subfigimg[width=\textwidth]{(c)}{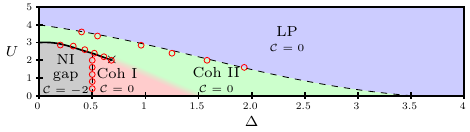}{8cm}{1.5}
\refstepcounter{subfigure}\label{fig:OppoPhaseDiagram}
\subfigimg[width=\textwidth]{(d)}{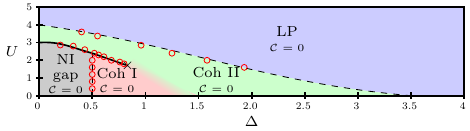}{8cm}{1.5} \refstepcounter{subfigure}\label{fig:SamePhaseDiagram}
  \end{subfigure}

  \end{subfigure}
\begin{subfigure}{0.49\linewidth}
    \centering
\subfigimg[width=0.8\textwidth]{(e)}{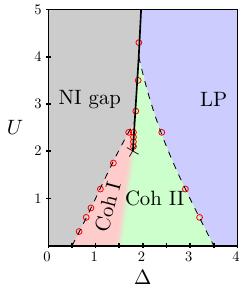}{1.6cm}{1.7}
\refstepcounter{subfigure}\label{fig:TwoBandPhaseDiagram}
\end{subfigure}%

\caption{(\subref{fig:U2pt2energy}) and (\subref{fig:U2pt0energy}): Energy states solving Hartree-Fock equations, demonstrating hysteresis for large $U$ which is absent for small $U$. (c) and (d): Phase diagrams for four-band models for crossing bands of opposite and the same Chern numbers, respectively.  (e): Phase diagram of a two-band model for crossing bands of opposite Chern numbers.}
\label{fig:5panelfig}
\end{figure*}

Specifically, we examine a two-component system with interactions such that inter-component coherent phases can be supported.  We demonstrate that the system supports different phases where the loops of maximum coherence in the Brillouin Zone are topologically distinct.  An example  of this is presented in Fig. \ref{fig:Lifshitz}.  The topologies of the Fermi surfaces on either side of the Lifshitz transition mirror the topologies of these maximum coherence loops, demonstrating that the Lifshitz transition ``seeds'' the quantum phase transition of the interacting system. Importantly, the two different coherent states are separated by a first order transition for relatively strong interactions. We find that, within the models we examine,  the transition line hosts a very unusual quantum critical endpoint (QCEP) \cite{RMP2016}, reminiscent of a thermodynamic $\mathcal{Z}_2$ critical point \cite{Chaikin1995,Pelissetto2002}.  As in the the latter, beyond the endpoint the evolution between coherent phases becomes continuous, with no sharp distinction between the states.  Examples of this are illustrated in Fig. \ref{fig:5panelfig}.

The phenomenon we describe in this work is robust with respect to band topology: we find it occurs whether the non-interacting bands that correlate through coherence have the same or different Chern numbers.  In the latter case, the first order transition can also entail a transition in the occupied band topology \cite{Amaricci2015PRL, Juricic2017}, but in such situations one does not find distinct coherent phases on either side of the transition.  Our study demonstrates that for systems where a non-interacting Lifshitz transition resides in a setting where spontaneous breaking of a continuous symmetry can occur, a rich set of transitions results, separating phases that offer insight into the different Fermi surface topologies that the non-interacting system supports.

\paragraph*{Model Hamiltonian.}\dash
We adopt a minimal model capturing the physics in which we are interested, with Hamiltonian of the form
\begin{equation}
\hat{H}=\sum_{\bf k} \sum_{\ell=t,b} \sum_{p=\pm 1} \left[pE_{\boldsymbol{k}}+\Delta_{\ell} \right] \hat{c}_{{\bf k},\ell,p}^{\dag}  \hat{c}_{{\bf k},\ell,p} + U \sum_{\bf k} \hat{\rho}_{-\boldsymbol{k},t} \hat{\rho}_{\boldsymbol{k},b},
\label{eq:Ham}
\end{equation}
where
$\hat{c}_{{\bf k},\ell,\pm 1}^{\dag}$ creates a particle in an eigenstate of a Bernevig-Hughes-Zhang Hamiltonian \cite{Bernevig_2006}
$h_{\ell,\boldsymbol{k}}$, with eigenvalues $\pm E_{\bf k}$ and eigenvectors $\chi_{{\bf q},\pm 1}$.
(Details may be found in the Supplementary Material (SM) \cite{SM}.) The second term is an interlayer contact interaction \cite{Lunde2013, Bozkurt2018} that involves density operators
$\hat{\rho}_{\boldsymbol{k},\ell}=\sum_{{\bf q},p=\pm 1} \chi^{\dag}_{{\bf k} +{\bf q},p} \cdot \chi_{{\bf q},p} \hat{c}_{{\bf q}+{\bf k},\ell,p}^{\dag}\hat{c}_{{\bf q},\ell,p}$.  Eq. \ref{eq:Ham} preserves the total number of fermions of $t$- and $b$-flavor separately, and supports a $U(1)$ flavor symmetry.  
Such models belong to the general class of extended Falicov-Kimball models (EFKM), which have been studied as candidates for describing electronic excitonic physics, ferroelectricity and bilayers phases \cite{Batista_2002,Freericks_2003,Batista_2004,Ihle2008,Phan2011,Golosov_2012,Zenker2012,Kaneko2013, Ejima2014,Fark_2023}. The impacts of Lifshitz transitions and band topology on such systems, the focus of this study, are to our knowledge unknown.

The model can also be interpreted as a bilayer system without tunneling between layers \cite{Zheng1997, Hanna2000, DasSarmaarXiv1, DasSarmaarXiv2}), in which intralayer interactions have been neglected.  In what follows we adopt this realization as a paradigm for such systems, and refer to the flavors as layers.  We expect the behavior of this model at half-filling to apply when bands of each flavor are relatively far apart energetically, while there is a crossing of bands of two different flavors: with short-range interactions, Fermi statistics suppresses short-range intra-flavor interactions, allowing inter-flavor interactions to dominate.  Beyond this, the model can also be mapped onto a two-species fermionic atomic gas system in an optical lattice, as we discuss below.

\paragraph*{Hartree-Fock Analysis.}\dash
We consider ground states of $\hat{H}$ within the Hartree-Fock (HF) approximation, in situations where the system is half-filled. Details of the analysis may be found in the SM \cite{SM}.  For $\Delta=0$, the spectrum consists of one filled and one empty band for each layer, with an intervening gap that is present even in the absence of interactions.  With increasing $|\Delta|$, bands of opposite flavor approach one another, eventually crossing when spontaneous symmetry breaking is not considered.  The Fermi surfaces consist of matching loops for each flavor surrounding the $\Gamma$ point of the square Brillouin zone (BZ).  With growing $|\Delta|$ these loops eventually touch the $M$ points at the BZ edge, signaling a Lifshitz transition.  For still larger $|\Delta|$ the loop topology changes, now surrounding the $X$ points (corners) of the BZ.

Multi-flavor systems with matching Fermi surfaces are known to be unstable to spontaneous inter-flavor coherence in the presence of interactions \cite{Min2008, Seradjeh2009}.  Figs. \ref{fig:5panelfig} (c) and (d) illustrates what happens, within our HF analysis, when two bands pass fully through one another as a function of $\Delta$, with a Lifshitz transition separating different Fermi surface topologies (see SM \cite{SM} for an illustration.)  We consider both crossing bands of the same Chern number $\mathcal C=1$ [panel (c)], and crossing bands of opposite Chern number $\mathcal C= \pm 1$ [panel (d)]. Continuous transitions are indicated as dashed lines while solid lines indicate first order transitions.  The resulting phases include a non-interacting gap (NIG) phase, which is continuously connected to the $\Delta=0$ non-interacting state; inter-layer coherent phases (Coh I and II); and a layer-polarized (LP) phase.

To characterize the polarization of a phase, we define a polarization function
$    P(\boldsymbol{k}) \equiv \sum_{i=1}^4 \langle \psi_i(\boldsymbol{k}) | \tau_z \otimes \mathbbm{1} | \psi_i(\boldsymbol{k}) \rangle f(E_i(\boldsymbol{k}))
    \equiv \sum_{i=1}^4 P_i (\boldsymbol{k}) f(E_i(\boldsymbol{k})),
    $
where $|\psi_i({\bf k}) \rangle$ are the four HF wavefunctions at wavevector ${\bf k}$ with energy $E_i({\bf k})$, $f$ is the Fermi function, and the $\tau_z$ is a Pauli matrix acting in layer-space. Spontaneous coherence in the system arises when order parameters of the form $\langle c^{\dag}_{{\bf k},\ell,p} c_{{\bf k},\ell',p'} \rangle \ne 0$ for $\ell \ne \ell'$.  This always entails values of $P_i({\bf k})$ which are not equal to either $-1$ or $1$, so that non-vanishing values of a coherence function,
$
        C(\boldsymbol{k}) \equiv \sum_{\boldsymbol{k}} \sum_{i=1}^4 (1-|P_i(\boldsymbol{k})|) f(E_i(\boldsymbol{k})),
$
signal the presence of interlayer coherence.

\begin{figure}[htbp]
    \centering
    \includegraphics[width=\linewidth]{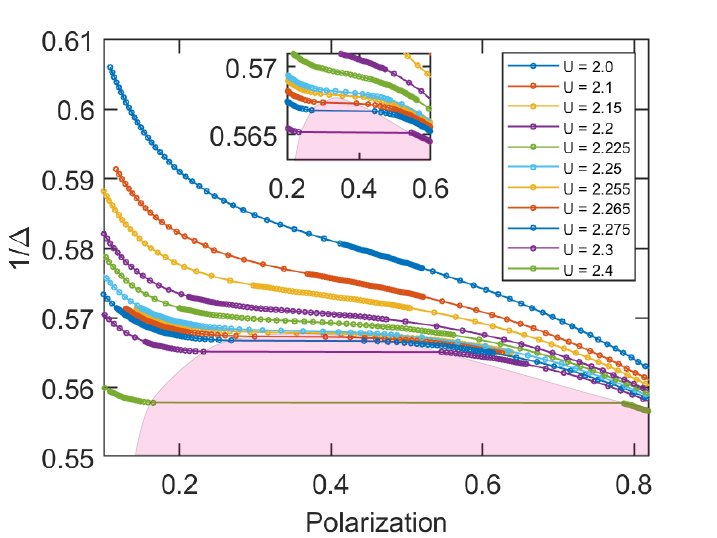}
    \caption{Plot of $P$ versus $1/\Delta$ for fixed $U$, illustrating a closing discontinuous polarization jump.}
    \label{fig:Amaricci}
\end{figure}

Fig. \ref{fig:Lifshitz} illustrates the behavior of $C({\bf k})$ as a function of ${\bf k}$ for two coherent states, and their correlation with the non-interacting Fermi surface.    The loci of maximum $C({\bf k})$ have two different topologies reflecting the behavior of the Fermi surfaces on either side of the non-interacting Lifshitz transition.  For small $U$ the evolution from one behavior to the other as a function of $\Delta$ is continuous, while for larger values there is a first order transition between them (see Figs. \ref{fig:5panelfig} (a) and (b)).  The transition line is quite interesting.  It hosts a very unusual QCEP, which at the mean-field level is highly analogous to the critical point of a thermal $\mathcal{Z}_2$ transition. Indeed, the behavior of the full polarization, $P = \sum_{{\bf k}}P({\bf k})$, when the bias $\Delta$ is varied, displays a jump that continuously vanishes at the QCEP, as illustrated in  Fig. \ref{fig:Amaricci}.

\begin{figure}[htb]
      \centering
\begin{subfigure}{0.49\linewidth}
     \subfigimg[width=\linewidth]{(a)}{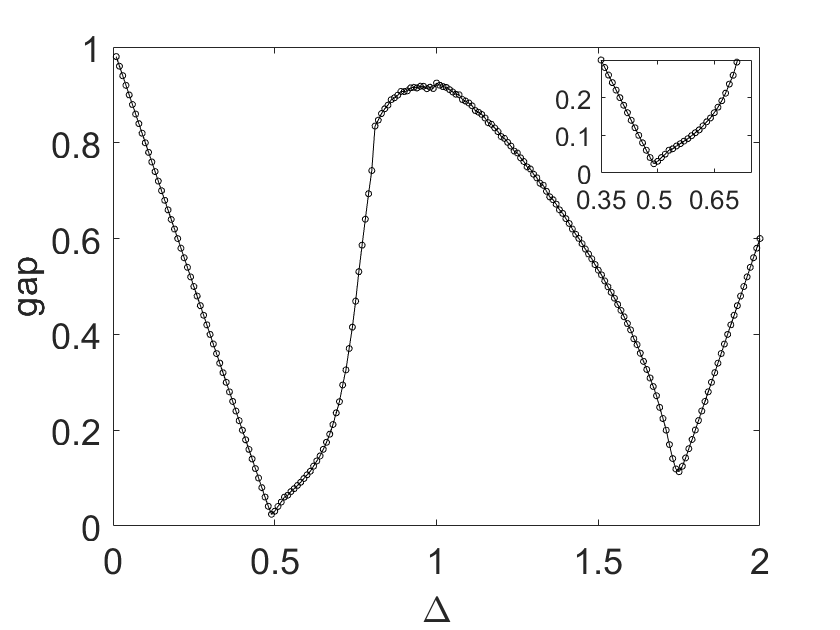}{0.8cm}{1.6}
     \refstepcounter{subfigure}\label{fig:SameGap}
  \end{subfigure}
      \hfill
\begin{subfigure}{0.49\linewidth}
     \subfigimg[width=\linewidth]{(b)}{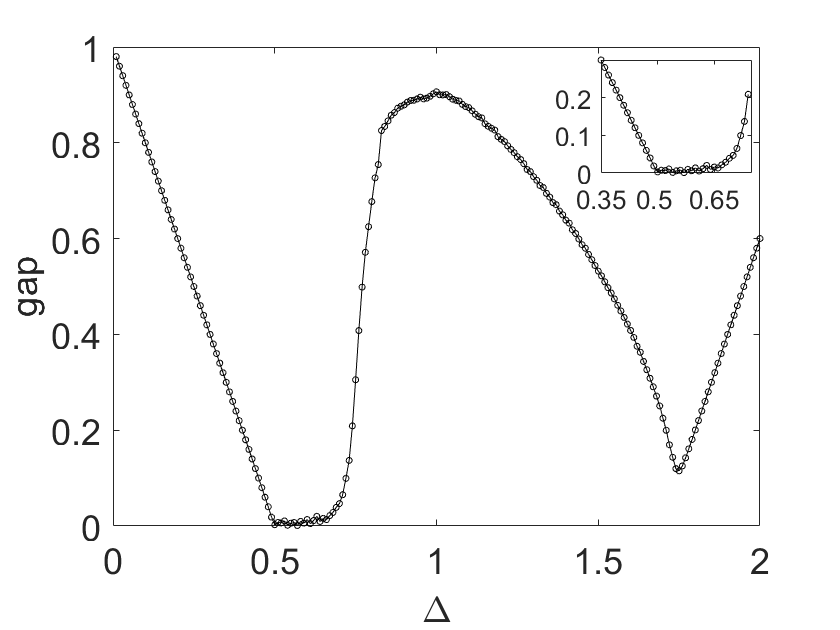}{0.8cm}{1.6}
     \refstepcounter{subfigure}\label{fig:OppoGap}
  \end{subfigure}
         \caption{Gap vs. bias $\Delta$ for $U =1.8$. (\subref{fig:SameGap}): Crossing bands have the same Chern number $\mathcal{C} = 1$. (\subref{fig:OppoGap}): Crossing bands have opposite Chern number $\mathcal{C} = \pm 1$. Insets: Gap remains open in first case but closes in second.}
         \label{fig:Gap vs bias}
 \end{figure}

Beyond this, as $U$ drops from large values, the transition line gives birth to two continuous transitions, in which coherence sets in from either the NIG or the LP phase.  Moreover, although the situations illustrated in Figs. \ref{fig:5panelfig} (c) and (d) look very similar, they have an important difference.  In (\subref{fig:OppoPhaseDiagram}), the transition out of the NIG phase involves a change in Chern number.  The transition occurs without a gap closing at large $U$ \cite{Amaricci2015PRL, Ezawa2013, Yang2013, Rachel2016}, while at smaller $U$ the onset of coherence occurs continuously and simultaneously with the topological transition.  The {\it topological} transition is in this way tied up with the {\it symmetry-breaking} transition.  By contrast, in (\subref{fig:SamePhaseDiagram}) there is no topological transition of the filled bands, and there is no closing of the band gap at any of the quantum phase transitions.  Fig. \ref{fig:Gap vs bias} illustrates this difference in behavior.

\paragraph*{Two Band Model.}\dash Several of the results described above can be qualitatively understood within a two band model, in which one retains only the two bands that cross one another in the model.  These can have the same or opposite topologies;  Fig. \ref{fig:5panelfig}(e) illustrates the resulting phase diagram for the case of opposite Chern numbers.  Although there is a change in the locations of the phase boundaries, particularly at large $U$, the system retains the same basic phases and features of the four band model.  Most prominent is the first order transition line dropping from large $U$, which ends at a QCEP.

The phases of this model are characterized by two order parameters: (i) $B_p\equiv {U \over 2 V}\sum_{\bf k}\langle \hat{c}^{\dag}_{{\bf k},t} \hat{c}_{{\bf k},t}-\hat{c}^{\dag}_{{\bf k},b} \hat{c}_{{\bf k},b} \rangle$, where $\hat{c}_{{\bf k},\ell}$ annihilates a particle in the retained band of layer $\ell$, which is a measure of the polarization of the system; and (ii) an interlayer coherence $B^{tb}({\bf k}) \equiv {U \over V} \sum_{{\bf k}_1} |\chi_{{\bf k},t}^{\dag} \cdot \chi_{{\bf k}_1,b}|^2\langle c^{\dag}_{{\bf k}_1,t} c_{{\bf k}_1,b} \rangle$.  ($V$ is the system area).  Within mean-field theory the self-consistent equations for these have the form \cite{SM}
\begin{align}
    \frac{B_p}{U} = \dfrac{1}{2V} \sum_{{\bf k}_2} &\dfrac{\widetilde{\xi}({\bf k}_2)}{\sqrt{|\widetilde{\xi}({\bf k}_2)|^2 + |B^{tb}({\bf k}_2)|^2}}
     \equiv F(B_p + \Delta);
     \label{eqn:bp}
     \\
    B^{tb}({\bf k}_1) = {U \over {2V}} \sum_{{\bf k}_2}
&\dfrac{f(\theta_{{\bf k}_1},\theta_{{\bf k}_1};\mathcal{C}_{rel})\, B^{tb}({\bf k}_2)}{\sqrt{|\widetilde{\xi}({\bf k}_2)|^2 + |B^{tb}({\bf k}_2)|^2}} , \label{eqn:btb}
\end{align}
where $\tilde{\xi}({\bf k}_2) = E_{{\bf k}_2} + \Delta+B_p$, $\mathcal{C}_{rel}=1(-1)$ when the product of the Chern numbers for the crossing bands is 1(-1), $f(\theta_{{\bf k}_1},\theta_{{\bf k}_1};1)=(1+ \cos\theta_{{\bf k}_1} \cos\theta_{{\bf k}_2})/{2}$ and
$f(\theta_{{\bf k}_1},\theta_{{\bf k}_1};-1)=\left(\cos^2\theta_{{\bf k}_1}/2 \right)\left(\cos^2\theta_{{\bf k}_2}/2\right)  $.
In Eq. \ref{eqn:btb} we have assumed that $\langle c^{\dag}_{{\bf k}_1,t} c_{{\bf k}_1,b} \rangle$ is real and has $C_4$ rotational symmetry in the HF ground state, which we indeed find in our more general numerical analysis.

Eq. \ref{eqn:bp} provides particular insight into the connection between the first-order transition lines in Fig. \ref{fig:5panelfig} and a thermal $\mathcal{Z}_2$ transition.  Direct plots of the left- and right-hand sides of the equation show that at large positive (negative) values of $\Delta$, one finds a single solution with maximal negative (positive) values of $B_p$, but in some transition region of $\Delta$ there are three solutions.  (See SM for details \cite{SM}.)  The physical state of the system jumps between two of the three solutions when their energies cross, behavior which is highly reminiscent of what one finds in a mean-field treatment of, for example, the liquid-gas transition \cite{Pathria4thEd}.  Such a jump always occurs provided $\max_x \frac{dF(x)}{dx} > \frac{1}{U}$.  Interestingly, because of the Lifshitz transition, one finds a point for which the left-hand side of the inequality {\it diverges} when $B^{tb}=0$, so that there is a first order jump for {\it any} positive value of $U$, and no QCEP is manifested.  However, for $|B^{tb}| > 0$, this divergence is smoothed over, $\max_x \frac{dF(x)}{dx} < \infty$, and for sufficiently small (but non-vanishing) $U$ the first order jump gives way to a continuous cross-over: a QCEP is stabilized.  This mean-field phenomenology is highly reminiscent of that of a classical liquid-gas critical point. Remarkably, in order to be realized in this setting, spontaneous coherence, a purely quantum phenomenon, must be manifested between layers.

Eq. \ref{eqn:btb} also allows an understanding of the behaviors of the coherence onset at the NIG-Coh I boundary for small $U$, which are different depending on whether $C_{rel}= \pm 1$.  In this regime, the bias is such that the single-particle energies of the two bands are very close near the $\Gamma$ point. Writing $\Delta_0$ for the bias at which the two bands touch at the $\Gamma$ point, we define $\tilde{\Delta} \equiv \Delta -\Delta_0$, and simplify our model by assuming the bands to be quadratic, in which case they have a constant density of states $g_0$.  Within this model and for small $k$ one may assume $\cos\theta_{{\bf k}} = \cos\theta(E_{\bf k}) \approx -1 + \alpha (E_{\bf k}-E_0)$, with $\alpha$ a parameter of order the bandwidth $W$.  As shown in the SM \cite{SM}, one may then show
$B^{tb} \approx b_0(1-\cos\theta_{\bf k})$ for $\mathcal{C}_{rel}=1$, and $B^{tb} \approx b_0\cos^2 \theta_{\bf k}/2$ for $\mathcal{C}_{rel}=-1$, with
\begin{align}
b_0 \approx  W &\exp \left[ - \frac{2}{Ug_0} \right] \quad\quad\quad \mathcal{C}_{rel}=1; \nonumber \\
b_0 \approx 4 \sqrt{\frac{W}{\alpha^2 |\tilde{\Delta}|}} &\exp \left[ \frac{-4}{\alpha^2 \tilde{\Delta}^2 U g_0} \right] \Theta(-\tilde{\Delta}) \quad \mathcal{C}_{rel}=-1;\nonumber
\end{align}
for small $|\tilde{\Delta}|$.  In these expressions, $\Theta$ is a Heaviside step function.
These results show that for bands of opposite Chern number ($\mathcal{C}_{rel}=-1$) the band closing ($\tilde{\Delta} \rightarrow 0$) associated with a change of topology of the occupied band occurs simultaneously with the onset of coherence ($b_0 \ne 0$).  This contrasts with the situation for $\mathcal{C}_{rel}=1$, in which $b_0$ is non-zero for small $|\tilde{\Delta}|$ even if the corresponding non-interacting bands do not cross.  Thus the single particle energy gap never closes in this case.     The differing Berry's curvatures of the two bands for $\mathcal{C}_{rel}=-1$ case frustrates the formation of coherence in the system.  This is apparent in the insets of Fig. \ref{fig:Gap vs bias}.

\paragraph*{Discussion.}\dash
These results suggest a number of interesting questions.
One set of these addresses the universality classes of the critical point as well as those of the coherence onset regions when the bands first cross.  In the former case, the presence of a broken $U(1)$ symmetry and its accompanying Goldstone mode suggests that its critical behavior will be different than that of a classical thermal $\mathcal{Z}_2$ transition.  In the latter case, the presence of gapless fermions for $\mathcal{C}_{rel} = -1$ suggests the transition will be in a different universality class than for $\mathcal{C}_{rel} = 1$.
Effects of real thermal fluctuations on the system, and the form a quantum critical region \cite{Sachdev2011, RMP1997} takes, is import to understand in settings where temperature effects cannot be ignored.  Another set of questions involve how this phase diagram might be manifested in different physical realizations.  One involves an optical lattice \cite{Jotzu2014} hosting two species of atoms with an inter-species Feshbach resonance \cite{Dao_2012,Grobner2017}.  A particle-hole transformation maps this onto an EFKM; in this case coherence is realized in superconducting states of the system.  For appropriate parameters, we expect {\it two} such states, separated by a first order transition.  Finally, van der Waals materials \cite{PT2016} offer platforms for electron bilayer realizations of this system, which support layer polarized states \cite{Young2011,Hunt_2017} and/or interlayer coherence \cite{Moon_1995,Eisenstein_2003,Lutchyn_2010,Murthy_2017,Li_2019}, whose interaction and competition could lead to novel quantum phase boundaries and transitions such as those we have described in this study.

\paragraph*{Acknowledgements}\dash The authors acknowledge useful discussions with Chunli Huang, Ganpathy Murthy, and Phil Richerme. This research was supported in part by Lilly Endowment, Inc., through its support for the Indiana University Pervasive Technology Institute. 
This work is supported in part by NSF Grant Nos. DMR-1914451 and ECCS-1936406. The authors thank the Aspen Center for Physics (NSF Grant No.
PHY-1607611) where part of this work was done.

\bibliography{thebibliography}

\end{document}


\title{Supplemental Material for ``Impact of a Lifshitz Transition on the onset of spontaneous coherence"}

\author{Adam Eaton}
\affiliation{Department of Physics, Indiana University, Bloomington, Indiana 47405, USA}

\author{Dibya Mukherjee}
\affiliation{Department of Physics, Indiana University, Bloomington, Indiana 47405, USA}

\author{H. A. Fertig}
\affiliation{Department of Physics, Indiana University, Bloomington, Indiana 47405, USA}
\affiliation{Quantum Science and Engineering Center, Indiana University, Bloomington, Indiana 47405, USA}

\date{\today} 

\begin{abstract}
Here we provide details of our self-consistent equations and further details about the topological transition. For the self-consistent  equations, we show
the calculations of $B^{tb}$ (i.e., the coherence) and demonstrate how the results depend on the topology of the bands. We also provide further numerical and analytic analysis on the system's behavior near topological transitions.
\end{abstract}

\keywords{first keyword, second keyword, third keyword}

\maketitle

\onecolumngrid





\section{Hartree Fock Equations for Four Band EFKM}

The single particle Hamiltonian that we consider is based on the Bernevig-Hughes-Zhang model \cite{Bernevig_2006}, which supports nontrivial band topology \cite{Ren2020, Lunde2013, Seshadri2019, Liu2010, Asboth2016}. For electrons belonging to flavor $\ell = t, b$ and momentum ${\bf k}$ this Hamiltonian is
\begin{align}
h_{\ell,\boldsymbol{k}} &= s_\ell \left( \hbar v s_{k_x} \sigma_x + \hbar v s_{k_y} \sigma_y + M_{\ell, \boldsymbol{k}} \sigma_z \right) + \Delta_\ell \mathbb{1} , \label{eqn:ham}
\end{align}
where $s_\ell=\pm 1$ is an index used to control whether bands of the same or opposite Chern numbers 
 cross one another, $\Delta_\ell = \pm\Delta$ is a bias experienced by the $\ell$-type electrons, $s_{k_\mu}=\sin k_{\mu}$ with $\mu=x,y$ (our unit of length is the lattice constant), $\sigma_i$ $(\ i\in\{x,y,z\})$ are Pauli matrices acting on two internal orbitals associated with each tight-binding site, and $v$ is the Fermi velocity.  In units where $\hbar v=1$, $M_{\ell,\boldsymbol{k}} = m_\ell + 2 - \cos k_x - \cos k_y$.  For $-4 < m_\ell <0$, the two bands of each family have Chern number $\mathcal C = \pm 1$; outside this interval their Chern numbers are zero. For simplicity we take $m_\ell = -0.5$, so that $M_{\ell, \boldsymbol{k}} = M_{\boldsymbol{k}}$ is independent of the $\ell$ index, and focus on total fermion densities such that the system is at half-filling. Occupation of the two families of electrons is controlled by varying $\Delta$, and we focus on ranges of this parameter such that (in the absence of interactions) the lower energy band of one family crosses the upper energy band of the other. The eigenstates of the Hamiltonian given by Eq. \eqref{eqn:ham} are
\begin{align}
\chi_{\boldsymbol{k},+} =
\dfrac{1}{\sqrt{2E_{\boldsymbol{k}} (E_{\boldsymbol{k}} + M_{\boldsymbol{k}})}}
&\begin{pmatrix}
    E_{\boldsymbol{k}} + M_{\boldsymbol{k}} \\ s_{k_x} + is_{k_y} \\ \end{pmatrix}
    \equiv
\begin{pmatrix}
    \cos \theta_{\boldsymbol{k}}/2 \\ \sin \theta_{\boldsymbol{k}}/2 \ e^{i \varphi_{\boldsymbol{k}}} \\ \end{pmatrix} \nonumber \\
     \chi_{\boldsymbol{k},-}
    =
    &\begin{pmatrix}
    -\sin \theta_{\boldsymbol{k}}/2 \ e^{-i \varphi_{\boldsymbol{k}}}\\ \cos \theta_{\boldsymbol{k}}/2 \\\end{pmatrix} \label{eqn:wfs},
\end{align}
where $+$ and $-$ correspond to the positive and negative energy bands for $s_{\ell}=1$ and $\Delta_\ell=0$, and
$E_{\boldsymbol{k}}= \sqrt{s_{k_x}^2+s_{k_y}^2+M_{\boldsymbol{k}}^2}$. 
For $s_{\ell}=-1$, the wavefunctions $\chi_{{\bf k},+}$ and $\chi_{{\bf k},-}$ are interchanged.

To this single particle Hamiltonian we add an interaction of the Falicov-Kimball type, so that
the extended Falicov-Kimball model (EFKM) we work with has the form
\begin{equation}
\hat{H}=\sum_{\bf k} \sum_{\ell=t,b} \sum_{p=\pm 1} \left[pE_{\boldsymbol{k}}+\Delta_{\ell} \right] \hat{c}_{{\bf k},\ell,p}^{\dag}  \hat{c}_{{\bf k},\ell,p} + U \sum_{\bf k} \hat{\rho}_{-\boldsymbol{k},t} \hat{\rho}_{\boldsymbol{k},b},
\label{eq:Ham}
\end{equation}
where $\Delta_t=\Delta$, $\Delta_b=-\Delta$, and $\hat{\rho}_{\boldsymbol{k},\ell}=\sum_{{\bf q},p=\pm 1} \chi^{\dag}_{{\bf k} +{\bf q},p} \cdot \chi_{{\bf q},p} \hat{c}_{{\bf q}+{\bf k},\ell,p}^{\dag}\hat{c}_{{\bf q},\ell,p}$.  $U$ is an inter-flavor contact interaction, and we consider only repulsive interactions ($U \ge 0$.) The last term has the explicit form
\begin{align}
    H_{\text{inter}} &\equiv U \sum_{\bf k} \hat{\rho}_{-\boldsymbol{k},t} \hat{\rho}_{\boldsymbol{k},b} \nonumber\\
    &= \sum_{\bf k} \sum_{{\bf q}_1,{\bf q}_2} \sum_{p_1,p_2,p_3,p_4=\pm 1}
    G_{p_1,p_2,p_3,p_4}({-\bf k}+{\bf q}_1,{\bf q}_1,{\bf k}+{\bf q}_2,{\bf q}_2)
    \hat{c}_{{\bf q}_1-{\bf k},t,p_1}^{\dag}\hat{c}_{{\bf q}_1,t,p_2}
    \hat{c}_{{\bf q}_2+{\bf k},b,p_3}^{\dag}\hat{c}_{{\bf q}_2,b,p_4},
\end{align}
where
$$
G_{p_1,p_2,p_3,p_4}({\bf q}_1,{\bf q}_2,{\bf q}_3,{\bf q}_4)={U \over V}\left[\chi^{\dag}_{{\bf q}_1,p_1} \cdot \chi_{{\bf q}_2,p_2}\right]
    \left[\chi^{\dag}_{{\bf q}_3,p_3} \cdot \chi_{{\bf q}_4,p_4}\right].
$$
Here $V$ is the area of the system.  
We proceed by performing a Hartree-Fock decomposition on $H_{\text{inter}}$ to obtain an effective, mean-field single-particle Hamiltonian.  For this we need expectation values $\langle \hat{c}_{{\bf q},\ell,p}^{\dag}\hat{c}_{{\bf q}',\ell',p'}\rangle \equiv \eta_{p,p'}^{\ell,\ell'}({\bf q} )\delta_{{\bf q},{\bf q}'}$.  Note by considering only states that are diagonal in wavevector, we rule out states with in-plane charge ordering.  This is natural for half-filling in these systems.  We thus make the substitution $\hat{H}_{\rm inter} \rightarrow \hat{H}_{\rm inter}^{HF}$, in which
\begin{align}
\hat{H}_{\text {inter }}^{HF}= & \sum_{{\bf k}} \sum_{p,p^{\prime}} \Bigl[\, B_{p p^{\prime}}^{t t}\left({\bf k} \right) \hat{c}_{{\bf k},t, p}^{\dagger} \hat{c}_{{\bf k},t,p^{\prime}}+B_{p p^{\prime}}^{b b}\left({\bf k}\right) \hat{c}_{{\bf k},b,p}^{\dagger} \hat{c}_{{\bf k},b,p^{\prime}} \nonumber\\
& +B_{p p^{\prime}}^{tb}\left({\bf k}\right) \hat{c}_{{\bf k},t,p}^{\dagger} \hat{c}_{{\bf k},b,p^{\prime}}+B_{p p^{\prime}}^{b t}\left({\bf k}\right) \hat{c}_{{\bf k},b,p}^{\dagger} \hat{c}_{{\bf k},t,p^{\prime}} \Bigr],
\label{eq:H_inter_HF}
\end{align}
where
\begin{align}
    &B_{p p^{\prime}}^{t t}\left({\bf k}\right)=\sum_{{\bf k}_2} \sum_{p_3,p_4} G_{p,p^{\prime},p_3,p_4}\left({\bf k}, {\bf k},{\bf k}_2,{\bf k}_2\right) \eta_{p_3,p_4}^{b b}\left({\bf k}_2\right) , \label{eq:Btt}\\
    &B_{p p^{\prime}}^{b b}\left({\bf k}\right)=\sum_{{\bf k}_1} \sum_{p_1,p_2} G_{p_1,p_2,p,p'}\left({\bf k}_1, {\bf k}_1,{\bf k},{\bf k}\right) \eta_{p_1,p_2}^{t t}\left({\bf k}_1\right) , \label{eq:Bbb}\\
    &B_{p p^{\prime}}^{t b}\left({\bf k}\right)=-\sum_{{\bf k}_1} \sum_{p_2,p_3} G_{p,p_2,p_3,p'}\left({\bf k}, {\bf k}_1,{\bf k}_1,{\bf k}\right) \eta_{p_3,p_2}^{b t}\left({\bf k}_1\right) , \label{eq:Btb}\\
    &B_{p p^{\prime}}^{b t}\left({\bf k}\right)=-\sum_{{\bf k}_2} \sum_{p_1,p_4} G_{p_1,p',p,p_4}\left({\bf k}_2, {\bf k},{\bf k},{\bf k}_2\right) \eta_{p_1,p_4}^{t b}\left({\bf k}_2\right) . \label{eq:Bbt}
\end{align}
Our Hartree-Fock Hamiltonian is then
$$
\hat{H}^{HF}=\sum_{\bf k} \sum_{\ell=t,b} \sum_{p=\pm 1} \left[pE_{\boldsymbol{k}}+\Delta_{\ell} \right] \hat{c}_{{\bf k},\ell,p}^{\dag}  \hat{c}_{{\bf k},\ell,p}
+
\hat{H}_{\text {inter }}^{HF}.
$$
The Hartree-Fock approximation in this system consists of diagonalizing the effective single-particle Hamiltonian $\hat{H}^{HF}$, and from this computing the quantities $\eta_{p,p'}^{\ell,\ell'}({\bf q} )$ assuming that the ground state can be represented by filling half of the Hartree-Fock eigenstates, specifically those with the lowest eigenvalues of $\hat{H}^{HF}$.  Since the states and energies depend on $\eta$, the procedure is carried out iteratively until the values of $\eta$ used as input for $\hat{H}^{HF}$ are the same as those produced by its eigenstates.

Our numerical procedure was performed using a square grid of momentum points.  For some initial values of $(U, \Delta)$, all elements of $B^{ij}$ were typically set equal to an initial seed of all ones. The choice of the seed does not appear to impact the results of the HF calculation provided that the values of $(U, \Delta)$ are sufficiently far away from a first-order phase boundary. In a few cases, we introduced an element of randomness to the seed to verify numerical stability. The evaluation of these quantities was iterated until the difference between the values of $B^{ij}$ in two successive iterations was less than a fixed convergence criterion. After a solution was found for a set of parameters, this solution was typically used as a seed for parameters $(U, \Delta + \delta \Delta)$, producing results behaving as in either Fig. 2(a) or Fig. 2(b) of the main text. We used a typical grid size of $50 \times 50$, and usually results were considered converged when the relative change of the $B^{ij}$'s was less than $0.005$.
For many parameter sets we checked that results on a scale that would be visible in our figures did not change when finer grids or a more stringent convergence criterion was used.  For results near phase boundaries, this was sometimes necessary.

\section{Relation between Single Particle Band Structures and Phases}

\begin{figure}[hbtp]
  \centering
\begin{subfigure}{0.31\linewidth}
     \subfigimg[width=\linewidth]{\textcolor{black}{(a)}}{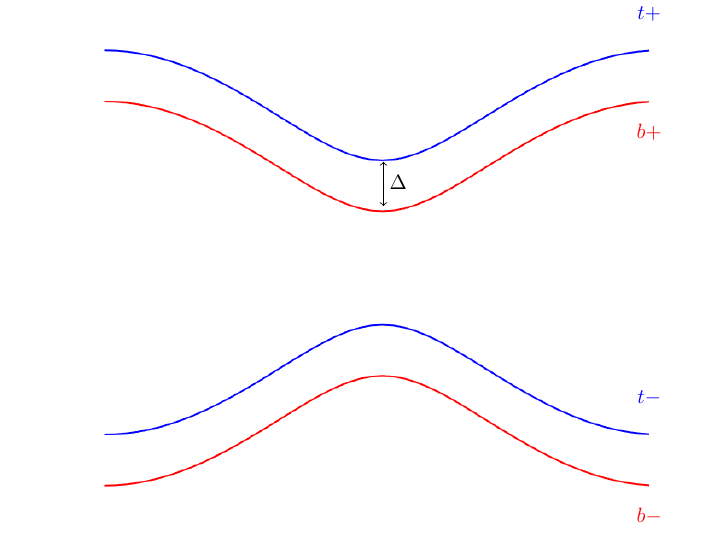}{0.2cm}{1.5}
     \refstepcounter{subfigure}\label{fig:NIG}
  \end{subfigure}
  \begin{subfigure}{0.31\linewidth}
    \subfigimg[width=\linewidth]{\textcolor{black}{(b)}}{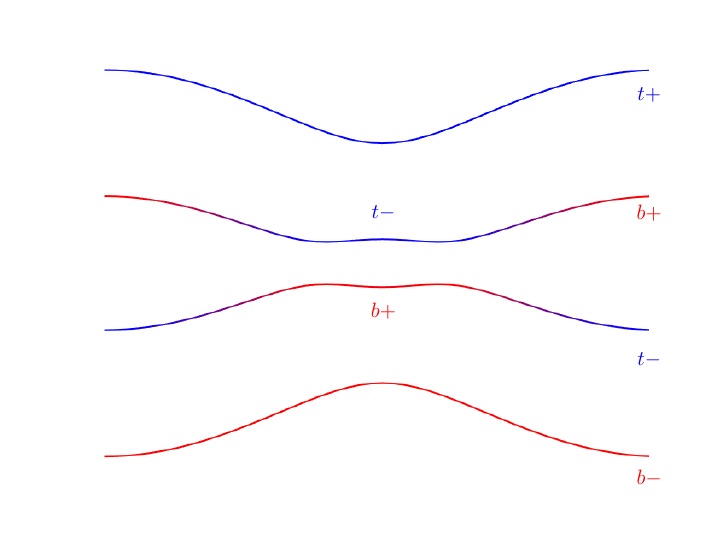}{0.2cm}{1.5}
    \refstepcounter{subfigure}\label{fig:Coh}
  \end{subfigure}
    \begin{subfigure}{0.31\linewidth}
    \subfigimg[width=\linewidth]{\textcolor{black}{(c)}}{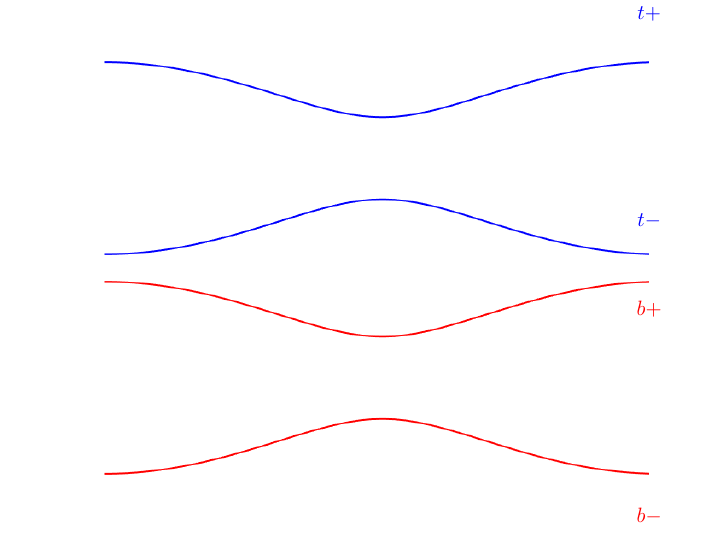}{0.2cm}{1.5}
    \refstepcounter{subfigure}\label{fig:LP}
  \end{subfigure}
  \caption{Band diagram sketches of the different phases exhibited by our model. Panel (\subref{fig:NIG}) represents the NIG phase, panel (\subref{fig:Coh}) a coherent phase, and panel (\subref{fig:LP}) the LP phase.
}
  \label{fig:NIGvsCohvsLP}
\end{figure}

The different phases presented by our models can be distinguished by the qualitative forms of their band structures. Fig. S1(\subref{fig:NIG}) shows bands which have been slightly renormalized by the interlayer interaction, yet are qualitatively the same as in the non-interacting case, for vanishing or small interlayer bias.   We call this phase the ``non-interacting gap" phase (NIG in the main text), and it is always characterized by vanishing $B^{tb}$ and $B^{bt}$ order parameters. Fig. S1(\subref{fig:Coh}) shows a bandstructure where the two bands nearest the Fermi level intermix, resulting in a coherent phase. Fig. S1(\subref{fig:LP}) displays a situation in which the bands have separated entirely by layer, resulting in a layer-polarized phase (LP in the main text).  Again in this phase, $B^{tb}$ and $B^{bt}$ vanish.

\begin{figure}[hbtp]
  \centering
\begin{subfigure}{0.2\linewidth}
     \subfigimg[width=\linewidth]{\textcolor{black}{(a)}}{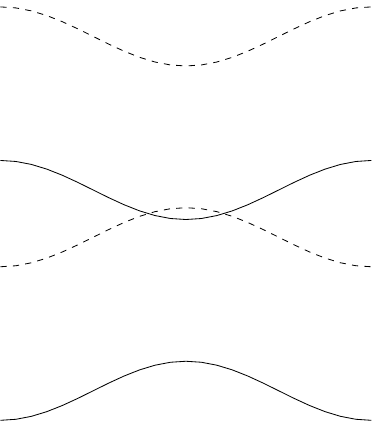}{0.2cm}{1.5}\vspace{1cm}
     \refstepcounter{subfigure}\label{fig:CohI}
  \end{subfigure}
  \begin{subfigure}{0.3\linewidth}
     \subfigimg[width=\linewidth]{\textcolor{black}{(b)}}{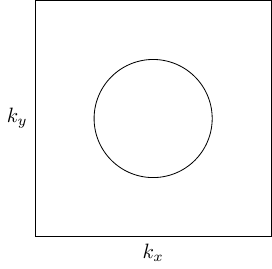}{0.85cm}{1.5}
     \refstepcounter{subfigure}\label{fig:FSI}
  \end{subfigure}
  \hfill \\
  \begin{subfigure}{0.2\linewidth}
    \subfigimg[width=\linewidth]{\textcolor{black}{(c)}}{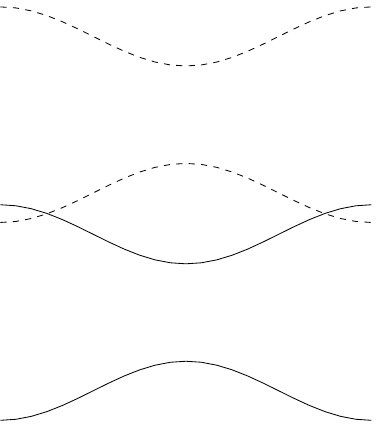}{0.2cm}{1.5}\vspace{1cm}
    \refstepcounter{subfigure}\label{fig:CohII}
  \end{subfigure}
    \begin{subfigure}{0.3\linewidth}
     \subfigimg[width=\linewidth]{\textcolor{black}{(d)}}{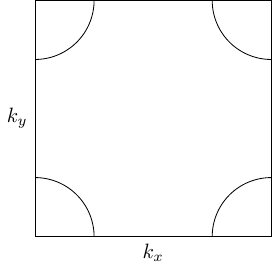}{0.85cm}{1.5}
     \refstepcounter{subfigure}\label{fig:FSII}
  \end{subfigure}
  \caption{{Band diagram sketches of the two different states that our coherent phases descend to when the order parameters $B^{tb}$ and $B^{bt}$ are set to zero by hand.} Panels (\subref{fig:CohI}) and (\subref{fig:CohII}) show the {non-coherent} bands, (\subref{fig:FSI}) and (\subref{fig:FSII}) are sketches of the Fermi surfaces corresponding to these band structures.}
  \label{fig:CohIvsCohII}
\end{figure}

Coherence in the system is manifested in two distinct ways. In one phase (Coh I) the bands cross near the $\Gamma$ point of the Brillouin zone (BZ), while in the other (Coh II) the bands cross near the $X$ point. Fig. S2(\subref{fig:CohI}) illustrates the situation for the first of these, and Fig. S2(\subref{fig:CohII}) illustrates the second. If one eliminates coherence between the two layers (setting $B^{tb}$ and $B^{bt}$ to zero in Eqs. \ref{eq:Btb} and \ref{eq:Bbt}), these phases adiabatically connect to states with Fermi surfaces illustrated in in Fig. S2(\subref{fig:FSI}) and Fig. S2(\subref{fig:FSII}), respectively.  These states are topologically distinct and are separated by a Lifshitz transition.  As illustrated in the main text, the different topologies are reflected in the coherent phases by the topologies of their loops of maximum coherence.

\newpage

\section{Self-Consistent Equations for Two-Band Model}

In the previous section we showed that all of the $B^{ij}_{p_1,p_2}$'s involved four different spinors.  States of the bands that actually anti-cross are in general linear combinations the four different eigenstates of $\hat{H}$ with $U=0$ with a given wavevector ${\bf k}$.  One may better understand our results by focusing on the just the two central bands in the model, projecting away the bands that are outermost in energy.  In this way each layer hosts a single band, so that the band index $p$ becomes extraneous, and the spinors for each layer have fixed form.  This allows the mean-field equations to be written in a relatively simple manner.

In the two-band model, in Eqs. \ref{eq:H_inter_HF} - \ref{eq:Bbt}, a single index $p$ is now associated with the layer index ($t$ or $b$), but these equations remain otherwise the same.  Using $\eta^{\ell,\ell'}_{p,p'}({\bf k}) \rightarrow \eta^{\ell,\ell'}({\bf k}) \equiv \langle \hat{c}_{{\bf k},\ell}^{\dag} \hat{c}_{{\bf k},\ell'}\rangle$ in Eqs. \ref{eq:Btt} - \ref{eq:Bbt}, one finds

\begin{align}
 B^{tt} &= \dfrac{U}{V} \sum_{{\bf k}_2} \left(\chi_{{\bf k}_1,t}^{\dag} \cdot \chi_{{\bf k}_1,t} \right) \left(\chi_{{\bf k}_2,b}^{\dag} \cdot \chi_{{\bf k}_2,b}\right) \langle \hat{c}_{{\bf k}_2,b}^{\dag} \hat{c}_{{\bf k}_2,b} \rangle
 =\dfrac{U}{2V} \sum_{{\bf k}_2} \left( 1- \dfrac{\widetilde{\xi}({\bf k}_2)}{\sqrt{|\widetilde{\xi}({\bf k}_2)|^2 + |B^{tb}({\bf k}_2)|^2}} \right)
\label{eq:2bnd_Btt}
\\
B^{bb} &= \dfrac{U}{V} \sum_{{\bf k}_2} \left(\chi_{{\bf k}_1,b}^{\dag} \cdot \chi_{{\bf k}_1,b} \right) \left(\chi_{{\bf k}_2,t}^{\dag} \cdot \chi_{{\bf k}_2,t}\right) \langle \hat{c}_{{\bf k}_2,t}^{\dag} \hat{c}_{{\bf k}_2,t} \rangle
 =\dfrac{U}{2V} \sum_{{\bf k}_2} \left( 1+ \dfrac{\widetilde{\xi}({\bf k}_2)}{\sqrt{|\widetilde{\xi}({\bf k}_2)|^2 + |B^{tb}({\bf k}_2)|^2}} \right)
\label{eq:2bnd_Bbb}
\\
    B^{tb} ({\bf k}_1) &= \dfrac{U}{V} \sum_{{\bf k}_2} \left( \chi^\dag_{{\bf k}_1,t} \cdot \chi_{{\bf k}_2,t} \right)\left( \chi^\dag_{{\bf k}_2,b} \cdot \chi_{{\bf k}_1,b} \right)
    \langle \hat{c}_{{\bf k}_2,b}^{\dag} \hat{c}_{{\bf k}_2,t} \rangle \nonumber \\
    &=
    \dfrac{U}{2V} \sum_{{\bf k}_2} \left( \chi^\dag_{{\bf k}_1,t} \cdot \chi_{{\bf k}_2,t} \right)\left( \chi^\dag_{{\bf k}_2,b} \cdot \chi_{{\bf k}_1,b} \right)
    \dfrac{B^{tb} ({\bf k}_2)}{\sqrt{|\widetilde{\xi}({\bf k}_2)|^2 + |B^{tb}({\bf k}_2)|^2}}
    \label{eqn:Btb}
    \end{align}
where in the first two of these equations we have used the normalization $\chi_{{\bf k},\ell}^\dag \cdot \chi_{{\bf k},\ell}=1$, and $\tilde{\xi}({\bf k}_2) = E_{{\bf k}_2} + \Delta+B_p$.
 The self-consistent equation for
$B_p\equiv {U \over 2V}\sum_{\bf k}\langle c^{\dag}_{{\bf k},t} c_{{\bf k},t}-c^{\dag}_{{\bf k},b} c_{{\bf k},b} \rangle$
in the main text (Eq. 2) follows directly from the first two equations above.  While the full self-consistent equation for $B^{tb}({\bf k})$ allows it to have any fixed overall phase, reflecting that the solution has broken a $U(1)$ symmetry,
in writing Eq. \ref{eqn:Btb} we have assumed $B^{tb}({\bf k})$ to be real, so that $B^{tb}({\bf k})=B^{bt}({\bf k})$.


\subsection{Case 1: Crossing Bands of Opposite Chern number ($\mathcal{C}_{rel}=-1$)}

In this case our band wavefunctions are
\begin{align}
\chi_{{\bf k},t} &= \begin{pmatrix}
    \cos \theta_{\bf k}/2 \\ \sin \theta_{\bf k}/2 \ e^{i \varphi_{\bf k}} \\ \end{pmatrix} \\
     \chi_{{\bf k},b} &= \begin{pmatrix}
    -\sin \theta_{\bf k}/2 \ e^{-i \varphi_{\bf k}}\\ \cos \theta_{\bf k}/2 \\\end{pmatrix},
\end{align}
where $\cos \theta_{\bf k}/2$,  $\sin \theta_{\bf k}/2$ and $e^{i \varphi_{\bf k}}$ have the definitions given in Eq. \ref{eqn:wfs} above.
With the shorthand notation $\theta_{{\bf k}_1} \equiv \theta_{1}$, etc., one finds
\begin{align}
    &\chi^\dagger_{{\bf k}_1,t-} \cdot \chi_{{\bf k}_2,t-} \nonumber \\
    &= \sin \theta_1/2 \sin \theta/2 e^{i (\varphi_1 - \varphi_2)} + \cos \theta_1/2 \cos \theta_2/2
\end{align}
and
\begin{align}
    &\chi^\dagger_{{\bf k}_2,b} \cdot \chi_{{\bf k}_1,b} \nonumber \\
    &= \cos \theta_1/2 \cos \theta_2/2 + \sin \theta_1/2 \sin \theta_2/2 e^{i (\varphi_1 - \varphi_2)}.
\end{align}
This leads to
\begin{align}
    &\chi^\dagger_{{\bf k}_1,t} \cdot \chi_{{\bf k}_2,t} \chi^\dagger_{{\bf k}_2,b} \cdot \chi_{{\bf k}_1,b} \nonumber \\
    &= \left(\sin \theta_1/2 \sin \theta_2/2 \ e^{i (\varphi_1 - \varphi_2)} + \cos \theta_1/2 \cos \theta_2/2 \right)^2 \nonumber\\
    &= \cos^2 \theta_2/2 \  \cos^2 \theta_1/2 + \sin^2 \theta_1/2 \ \sin^2 \theta_2/2 \ e^{2i (\varphi_1 - \varphi_2)} \nonumber\\
    &\hspace{0.25cm}+2 \ \sin \theta_1 /2 \ \cos \theta_2 /2 \ \cos \theta_1 /2 \ \sin \theta_2 /2 \ e^{i (\varphi_1 - \varphi_2)} \nonumber
    \\
    &= {1 \over 4}  (1 + \cos \theta_2)(1 + \cos \theta_1) \nonumber\\
    &\hspace{0.25cm}+ {1 \over 4}  (1 - \cos \theta_1) (1 - \cos \theta_2) \ e^{2i (\varphi_1 - \varphi_2)} \nonumber\\
    &\hspace{0.25cm}+ {1 \over 2} \sin \theta_1 \ \sin \theta_2 \ e^{i (\varphi_1 - \varphi_2)} \label{eq:4chi_same}
\end{align}
We substitute Eq. \ref{eq:4chi_same} into Eq. \ref{eqn:Btb}, and then assume $\langle \hat{c}_{{\bf k}_2,b}^{\dag} \hat{c}_{{\bf k}_2,t} \rangle$ is invariant when ${\bf k}_2$ is rotated by $\pi/2$, which we have found to be the case in our four band model when starting from an asymmetric seed.  In this case the last two terms make no contribution to Eq. \ref{eqn:Btb} upon summation over ${\bf k}_2$, the resulting form for $B^{tb}$ is real, and $B^{bt}=B^{tb}$.
We then arrive at the equation
\begin{align}
    B^{tb} ({\bf k}_1)
    &= \dfrac{U}{8V} \sum_{{\bf k}_2} (1 + \cos \theta_{{\bf k}_1})(1+ \cos \theta_{{\bf k}_2}) \dfrac{B^{tb}({\bf k}_2)} {\sqrt{|\widetilde{\xi}({\bf k}_2)|^2 + |B^{tb}({\bf k}_2)|^2}} \nonumber\\
    &= \dfrac{U}{2V} \sum_{{\bf k}_2} \cos^2 \theta_{{\bf k}_1}/2\cos^2 \theta_{{\bf k}_2}/2 \dfrac{B^{tb}({\bf k}_2)} {\sqrt{|\widetilde{\xi}({\bf k}_2)|^2 + |B^{tb}({\bf k}_2)|^2}}
    \label{eqn:oppo}
\end{align}
which is the concrete form of Eq. 3 in the main text that we use in this case.



\subsection{Case 2: Crossing Bands with the Same Chern number ($\mathcal{C}_{rel}=1$)}

In this case we take
\begin{align}
    \chi_{{\bf k},t} = \chi_{{\bf k},b} &= \begin{pmatrix}
    \cos \theta_{\bf k}/2 \\ \sin \theta_{\bf k}/2 \ e^{i \varphi_{\bf k}} \\ \end{pmatrix},
\end{align}
With the same notation as in the previous case,
\begin{align}
    &\chi^\dagger_{{\bf k}_1,t} \cdot \chi_{{\bf k}_2,t} \nonumber \\
    &= \cos \theta_1/2 \cos \theta_2/2 + \sin \theta_1/2 \sin \theta_2/2 e^{i (\varphi_2 - \varphi_1)}
\end{align}
and
\begin{align}
    &\chi^\dagger_{{\bf k}_2,b} \cdot \chi_{{\bf k}_1,b} \nonumber \\
    &= \cos \theta_1/2 \cos \theta_2/2 + \sin \theta_1/2 \cos \theta_2/2 e^{i (\varphi_1 - \varphi_2)},
\end{align}
so that
\begin{align}
    &\chi^\dagger_{t-} (k_1) \cdot \chi_{t-}(k_2) \chi^\dagger_{b+}(k_2) \cdot \chi_{b+}(k_1) \nonumber \\
    &= \sin^2 \theta_2/2 \  \sin^2 \theta_1/2 + \cos^2 \theta_1/2 \ \cos^2 \theta_2/2 \nonumber\\
    &\hspace{0.25cm} +\sin \theta_1 /2 \ \cos \theta_2 /2 \ \cos \theta_1 /2 \ \sin \theta_2 /2 \ [e^{i (\varphi_1 - \varphi_2)} + e^{i (\varphi_2 - \varphi_1)}] \nonumber\\
    &= {1 \over 4} \ (1 - \cos \theta_2)(1 - \cos \theta_1) + {1 \over 4} \ (1 + \cos \theta_1) (1 + \cos \theta_2) \nonumber\\
    &\hspace{0.25cm}+ {1 \over 2} \sin \theta_1 \ \sin \theta_2 \cos (\varphi_2 - \varphi_1) \nonumber\\
    &= {1 \over 2} (1 + \cos \theta_1 \cos \theta_2 + \sin \theta_1 \ \sin \theta_2 \cos (\varphi_2 - \varphi_1))
\end{align}
Again dropping terms that vanish in Eq. \ref{eqn:Btb} upon summation over ${\bf k}_2$, we arrive at
\begin{align}
    B^{tb} ({\bf k}_1)
    &= \dfrac{U}{4V} \sum_{{\bf k}_2} (1 + \cos \theta_{{\bf k}_1} \cos \theta_{{\bf k}_2}) \dfrac{B^{tb} ({\bf k}_2)}{\sqrt{|\widetilde{\xi}({\bf k}_2)|^2 + |B^{tb}({\bf k}_2)|^2}} \label{eqn:same}
\end{align}
Collectively, this result and that of the previous case may be written in the form shown in Eq. 3 of the main text.


\section{Berry Flux Through the Transitions}
\begin{figure}[hbtp]
     \centering
     \begin{subfigure}{0.32\textwidth}
         \centering
\subfigimg[width=\textwidth]{(a)}{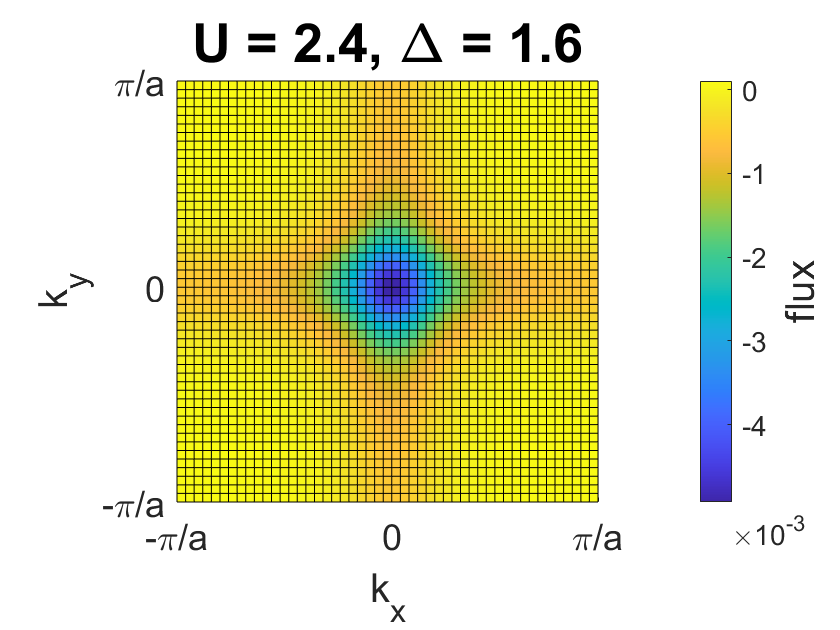}{1.3cm}{2.25}
         \refstepcounter{subfigure}\label{fig:1pt6}
     \end{subfigure}
     \hfill
     \begin{subfigure}{0.32\textwidth}
         \centering
\subfigimg[width=\textwidth]{(b)}{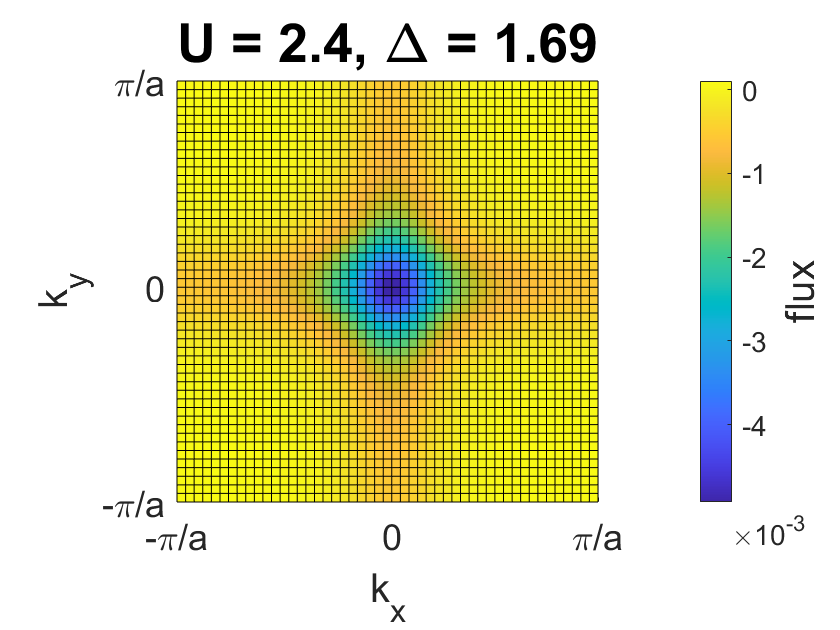}{1.3cm}{2.25}
         \refstepcounter{subfigure}\label{fig:1pt69}
     \end{subfigure}
     \hfill
     \begin{subfigure}{0.32\textwidth}
         \centering
\subfigimg[width=\textwidth]{\textcolor{white}{(c)}}{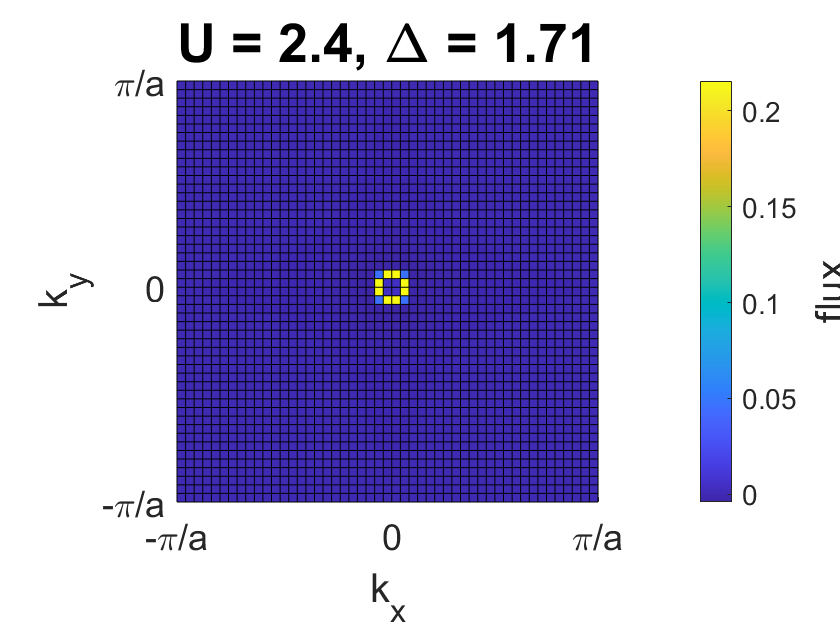}{1.3cm}{2.25}
         \refstepcounter{subfigure}\label{fig:1pt71}
     \end{subfigure}
     \hfill
     \begin{subfigure}{0.32\textwidth}
         \centering
\subfigimg[width=\textwidth]{\textcolor{white}{(d)}}{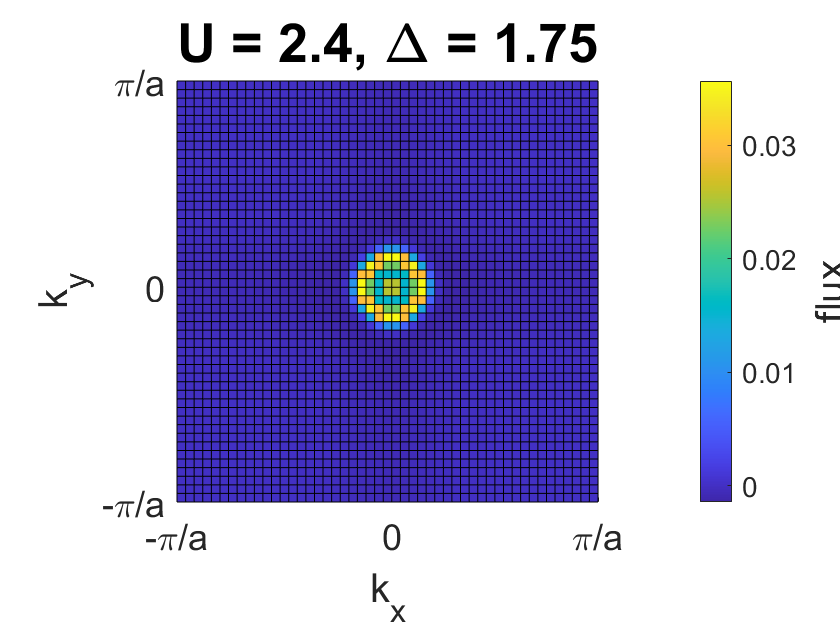}{1.3cm}{2.25}
         \refstepcounter{subfigure}\label{fig:1pt75}
     \end{subfigure}
     \hfill
          \begin{subfigure}{0.32\textwidth}
         \centering
\subfigimg[width=\textwidth]{\textcolor{white}{(e)}}{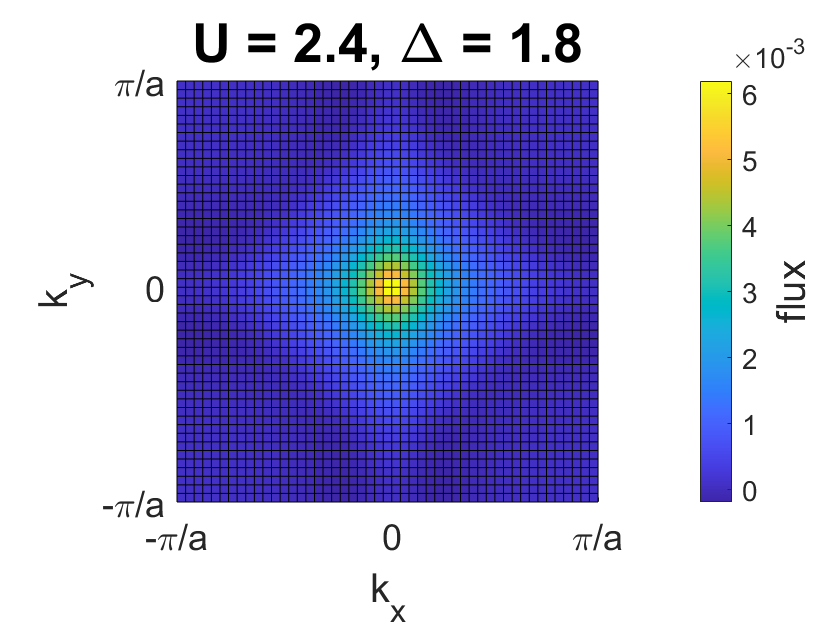}{1.3cm}{2.25}
         \refstepcounter{subfigure}\label{fig:1pt8}
     \end{subfigure}
     \hfill
     \begin{subfigure}{0.32\textwidth}
         \centering
\subfigimg[width=\textwidth]{\textcolor{white}{(f)}}{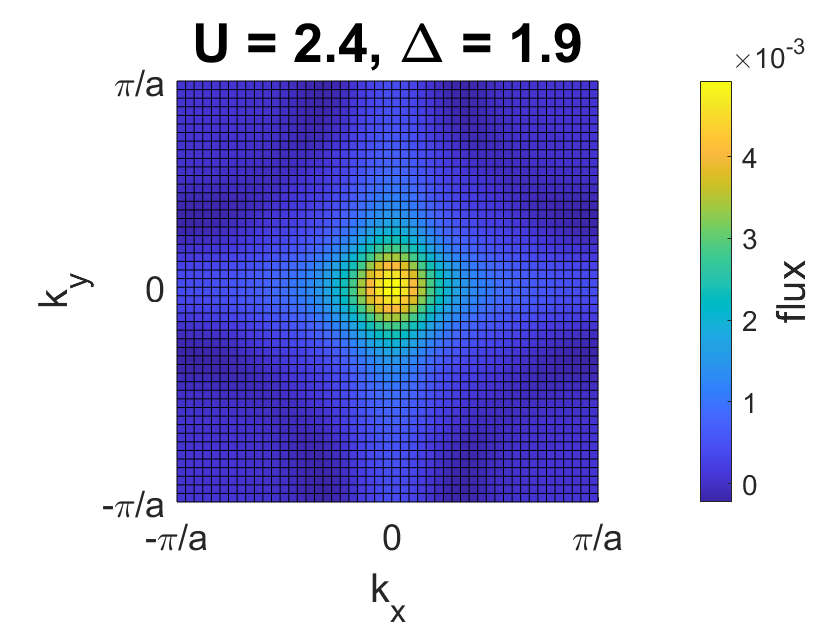}{1.3cm}{2.25}
         \refstepcounter{subfigure}\label{fig:1pt9}
     \end{subfigure}
     \caption{Evolution of the Berry's flux for fixed $U=2.4$ as a function of the bias $\Delta$. In panels (\subref{fig:1pt6}) and (\subref{fig:1pt69}) the Chern number of the occupied band is $\mathcal{C} = -1$. A topological transition accompanied by a gap closing occurs at $\Delta = 1.7$. In panels (\subref{fig:1pt71})-(\subref{fig:1pt9}) the occupied band has Chern number $\mathcal{C} = +1$.  (\subref{fig:1pt71}) and (\subref{fig:1pt75}) show results from the Coh I phase.  Between $\Delta = 1.75$ and $\Delta = 1.8$ there is a first-order transition into the Coh II phase. (\subref{fig:1pt8}) and (\subref{fig:1pt9}) show results from the Coh II phase.}
\label{fig:BerryFlux}
\end{figure}

In this section we present Berry flux results for the two band model as the system exits the NIG phase, passes through the Coh I phase, and finally transitions into the Coh II phase, for a situation where $\mathcal{C}_{rel}=-1$. For our choice of parameters, in the NIG phase the Berry's flux of the occupied band is negative and changes monotonically as one moves radially outward from $k=0$ [Figs. \ref{fig:BerryFlux} (a) and (b)]. Near the transition there is a pronounced minimum at the $\Gamma$ point. As the bias is increased and coherence forms, positive Berry's flux appears in the regions of the Brillouin zone exhibiting large coherence, and the Chern number of the occupied band changes from $-1$ to $+1$ [Figs. \ref{fig:BerryFlux}(c) and (d).]  With yet larger bias the system passes through the first order transition into the Coh II phase.  The Berry flux is considerably more spread out than in the Coh I phase, and the peak at the $\Gamma$ point is much smaller [Figs. \ref{fig:BerryFlux}(e) and (f).]

\newpage	

\section{Mean-Field Analysis of the Two-Band Model I: Origin of Critical Endpoint}
\begin{figure}[hbtp]
     \centering
     \begin{subfigure}{0.31\textwidth}
         \centering
         \subfigimg[width=\textwidth]{(a)}{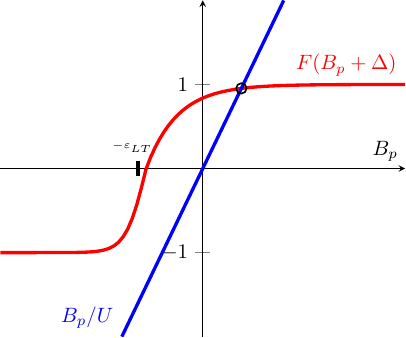}{0.4cm}{3.4}
         \refstepcounter{subfigure}\label{fig:MFcoha}
     \end{subfigure}
     \hfill
     \begin{subfigure}{0.31\textwidth}
         \centering
         \subfigimg[width=\textwidth]{(b)}{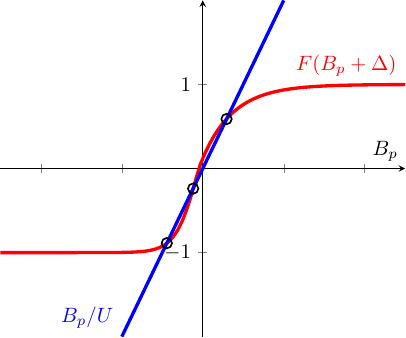}{0.4cm}{3.4}
         \refstepcounter{subfigure}\label{fig:MFcohb}
     \end{subfigure}
          \begin{subfigure}{0.31\textwidth}
         \centering
         \subfigimg[width=\textwidth]{(c)}{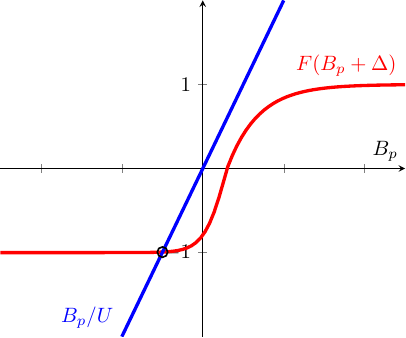}{0.4cm}{3.4}
         \refstepcounter{subfigure}\label{fig:MFcohc}
     \end{subfigure}
     \caption{Schematic showing the different possible solutions to Eq. \ref{eqn:bp_mft} for the polarization $B_p$. Panel (\subref{fig:MFcoha}) is at the largest bias,  panel (\subref{fig:MFcohb}) an intermediate bias, and panel (\subref{fig:MFcohc}) is at the smallest bias. In panel (\subref{fig:MFcoha}) $\varepsilon_{LT}$ refers to the $B_p$ at which the Lifshitz transition occurs when $B^{tb}=0$, for fixed $\Delta$. In the limit that $B^{tb} \to 0$, the slope of $F$ here becomes divergent, so that the transition from small to large $B_p$ solutions is first order for arbitrarily small $U$.  Finite values of $B^{tb}$ lower the maximum slope from this divergent value, allowing a QCEP to emerge in the phase diagram.}
     \label{fig:MFcoherenceAnalysis}
\end{figure}

The phases that we observe are characterized by two different order parameters -- the interlayer coherence $B^{tb}$ (discussed in more detail below) and the layer polarization $B_p$. The self-consistent equation for the latter has the form
\begin{align}
    \frac{B_p}{U} &= \dfrac{1}{2V} \sum_{{\bf k}_2} \left(\dfrac{\widetilde{\xi}({\bf k}_2)}{\sqrt{|\widetilde{\xi}({\bf k}_2)|^2 + |B^{tb}({\bf k}_2)|^2}} \right)
     \equiv F(B_p + \Delta).
     \label{eqn:bp_mft}
\end{align}
Depending on the values of $(U, \Delta)$, the equation above could have either one or three solutions. To see that this, we sketch the two sides of Eq. \eqref{eqn:bp_mft} in Fig. \ref{fig:MFcoherenceAnalysis} for different values of $\Delta$. The role of the bias in this picture is that increasingly negative values of $\Delta$ shift $F(B_p+\Delta)$ to the right. Moreover, decreasing $U$ increases the slope of $B_p/U$. Hence for extreme values of $\Delta$ (as in Figs. \ref{fig:MFcoha} and \ref{fig:MFcohc}) there will always be exactly one solution, while an intermediate bias (Fig. \ref{fig:MFcohb}) may support three solutions, provided $\max_x \frac{dF(x)}{dx} > \frac{1}{U}$, i.e., the maximum slope of $F$ must exceed the slope of $B_p/U$.  When three solutions are available, the two ``outer" solutions are local energy minima for the mean-field states, while the central solution is a local maximum. The first order transition occurs when the energy ordering of the states associated with the largest and smallest solutions for $B_p$ interchange.  Because the central solution merges with one of the outer solutions as $\Delta$ varies, this interchange of energy must occur as $\Delta$ is varied within the three solution range.

The roles of the Lifshitz transition and coherence are particularly interesting in this analysis.  In the incoherent limit $B^{tb} \to 0$, because the former induces a van Hove singularity in the density of states, the maximum slope of $F(B_p+\Delta)$ diverges when $\Delta$ is chosen such that the Fermi surface is precisely at the Lifshitz transition.
In this situation the first order transition will occur for arbitrarily small $U$, and the first order transition line ends on the $U=0$ axis precisely at the non-interacting Lifshitz transition.  There is no quantum critical endpoint (QCEP) in this situation.   For non-vanishing coherence ($B^{tb} \ne 0)$, the coherence ``smooths out"  the divergence in slope of $F(B_p+\Delta)$, yielding a maximum finite slope at $B_p \equiv -\varepsilon_{LT}$, indicated schematically in Fig \ref{fig:MFcoha}. Because of this, there exists a minimum $U$ below which the condition allowing three mean-field solutions ($\max_x \frac{dF(x)}{dx} > \frac{1}{U}$) cannot be satisfied. This means that as a function of $\Delta$, $B_p$ must evolve continuously from its minimum to its maximum value. This change in behavior indicates that the phase diagram supports a QCEP, and, at the mean-field level, this can {\it only} be present when the system supports inter-flavor coherence.

\newpage

\section{Mean-Field Analysis of the Two-Band Model II: Onset of Coherence at Small $U$}
\subsection{Case 1: Crossing Bands of the Same Chern number ($\mathcal{C}_{rel}=1$)}
Recall from Eq. \ref{eqn:same} that the self-consistent equation for $B^{tb}$ for the case where the two crossing bands have the same topology is
\begin{align}
    B^{tb} ({\bf k}_1)
    &= \dfrac{U}{4V} \sum_{{\bf k}_2} (1 + \cos \theta_{{\bf k}_1} \cos \theta_{{\bf k}_2}) \dfrac{B^{tb} ({\bf k}_2)}{\sqrt{|\widetilde{\xi}({\bf k}_2)|^2 + |B^{tb}({\bf k}_2)|^2}} \label{eqn:sameBtb}.
\end{align}
As a result, any self-consistent solution will be of the form $B^{tb}(\mathbf{k}) = b_0 + b_c \cos (\theta_{\mathbf{k}})$ where $b_0$ and $b_c$ are constants that do not depend on $\mathbf{k}$. As a simple model, we  assume that we can make the replacement $\theta_{\boldsymbol{k}} \rightarrow \theta (\varepsilon_{\bf k})$, where
$\varepsilon_{\bf k} = E_{\bf k} - E_{{\bf k}=0}$, so that we can rewrite Eq. \eqref{eqn:sameBtb} as two separate equations,
%
%
\begin{align}
b_{0}&=\left.\frac{U}{4} \int_{0}^{W} d \varepsilon g\left(\varepsilon\right) \frac{b_{0}+b_{c} \cos \theta\left(\varepsilon\right)}{\left[\left(\varepsilon+\widetilde{\Delta}\right)^{2}+\left(b_{0}+b_{c} \cos \theta\left(\varepsilon\right)\right)^{2}\right]^{1 / 2}}\right. , \\
b_{c}&=\left. \frac{U}{4} \int_{0}^{W} d \varepsilon g\left(\varepsilon\right) \cos \theta\left(\varepsilon\right) \frac{b_{0}+b_{c} \cos \theta\left(\varepsilon\right)}{\left[\left(\varepsilon+\widetilde{\Delta}\right)^{2}+\left(b_{0}+b_{c} \cos \theta\left(\varepsilon\right)\right)^{2}\right]^{1 / 2}}\right. .
\end{align}
Here $W$ is the bandwidth, $g(\varepsilon)$ is the density of states $g(\varepsilon)=\frac{1}{V} \sum_{\boldsymbol{k}} \delta\left[\varepsilon-\varepsilon_{\bf k}\right]$, and $\widetilde{\Delta} \equiv \frac{1}{2}\left[B^{tt}-B^{b b}\right]+\Delta$ is the renormalized bias.
%
%
%
%
%
%
%
Recalling that the wavefunctions for the BHZ Hamiltonian have the form (Eq. \eqref{eqn:wfs})
\begin{align}
\chi_{\boldsymbol{k},+} =
\dfrac{1}{\sqrt{2E_{\boldsymbol{k}} (E_{\boldsymbol{k}} + M_{\boldsymbol{k}})}}
&\begin{pmatrix}
    E_{\boldsymbol{k}} + M_{\boldsymbol{k}} \\ s_{k_x} + is_{k_y} \\ \end{pmatrix}
    \equiv
\begin{pmatrix}
    \cos \theta_{\boldsymbol{k}}/2 \\ \sin \theta_{\boldsymbol{k}}/2 \ e^{i \varphi_{\boldsymbol{k}}} \\ \end{pmatrix} \nonumber
    \end{align}
with $M_{\bf k}=m+2-\cos k_{x}-\operatorname{cosk}_{y}$ and $E_{\bf k}=\left(\sin ^{2} k_{x}+\sin ^{2} k_{y}+M_{\bf k}^{2}\right)^{1 / 2}$, we assume $\widetilde{\Delta}$ is small and negative, so that in the absence of coherence ($b_0=b_c=0$) the two bands cross very near $\varepsilon_{\bf k}=0$.  We then consider the behavior of $\cos\theta_{\bf k}$ near $k=0$.  In this limit,
$M_{\bf k} \approx m + \frac{k^2}{2}$ and $E_{\bf k} \approx |m|+\frac{(1+m)}{2 |m|} k^{2}$.
For small negative values of $m$ (recall we take $m=-1/2$ in our numerical analyses)
we expect that $\cos \theta_{{\bf k}=0} = -1$ exactly at the $\Gamma$ point, and for ${\bf k}$ slightly away from $\Gamma$,
\begin{align}
\cos \theta = \frac{M_{\bf k}}{E_{\bf k}}
&= \frac{-|m|+\frac{1}{2} k^{2}}{|m|+\frac{(1+m)}{2 |m|} k^{2}} \\
&\approx -1 + \frac{k^2}{2m^2}.
\end{align}
Since $E_{\bf k}-E_{{\bf k}=0}$ disperses quadratically from $k=0$, we make the replacement
$\cos \theta\left(\varepsilon\right) \to -1+\alpha \varepsilon$, where $\alpha$ is constant of order $1/W$.  This captures the behavior of the $\cos{\theta_{\bf k}}$ in the vicinity of the band crossing.

With this substitution, and taking the density of states near the band crossing to be constant ($g(\varepsilon) \approx g_0$), we arrive at the equations
%
%
%
\begin{align}
b_{0}&=\frac{U}{4} g_{0} \int_{0}^{W} d \varepsilon \frac{b_{0}+b_{c}[-1+\alpha \varepsilon]}{\left[\left(\varepsilon+\widetilde{\Delta}\right)^{2}+\left(b_{0}+b_{c}[-1+\alpha \varepsilon]\right)^{2}\right]^{1 / 2}} ,\\
b_{c}&= \frac{U}{4} g_{0} \int_{0}^{W} d \varepsilon(-1+\alpha \varepsilon) \frac{b_{0}+b_{c}[-1+\alpha \varepsilon]}{\left[\left(\varepsilon+\widetilde{\Delta}\right)^{2}+\left(b_{0}+b_{c}[-1+\alpha \varepsilon]\right)^{2}\right]^{1 / 2}} .
\end{align}

%
For small $U$ we expect $b_{0}$ and $b_{c}$ to be very small. The denominator then introduces a strong peak in the integrands of these equations near $\varepsilon=-\widetilde{\Delta}$, so we may write
\begin{align}
b_{c}&\approx \frac{U}{4} g_{0}\left[-1-\alpha\widetilde{\Delta}\right] \int_{0}^{W} d \varepsilon \frac{b_{0}+b_{c}[-1+\alpha \varepsilon]}{\left[\left(\varepsilon+\widetilde{\Delta}\right)^{2}+\left(b_{0}+b_{c}[-1+\alpha \varepsilon]\right)^{2}\right]^{1 / 2}}
\end{align}
Then $b_{c}=\left(-1-\alpha\widetilde{\Delta}\right) b_{0}$.  In the limit $|\widetilde{\Delta}|/W \ll 1$, we can set $b_c \approx -b_0$.  We then arrive at a single equation of the form
\begin{align}
1 &=\frac{U g_{0}}{4} \int_{0}^{W} d \varepsilon \frac{2-\alpha \varepsilon}{\left[\left(\varepsilon+\widetilde{\Delta}\right)^{2}+b_0^2\left(2-\alpha \varepsilon\right)^{2}\right]^{1 / 2}}
\end{align}
Anticipating that, for small $Ug_0$, $b_0$ will be very small we write the equation in form
$$
1=\frac{Ug_0}{4} \int_0^W d\varepsilon \frac{2-\alpha \varepsilon}{\sqrt{\left(\varepsilon +B\right)^2+C}},
$$
where we have defined the constants $B=\widetilde{\Delta}-2\alpha b_0^2$ and $C=\widetilde{\Delta}^2+4b_0^2-\left(\widetilde{\Delta} - 2\alpha b_0^2 \right)^2$, and have dropped a term of order $\alpha^2 b_0^2$.  Carrying through the integration yields
\begin{align}
1=\frac{Ug_0}{4}\Biggl\{ \left(2+\alpha B\right) &\left[\ln \left|\frac{W+B}{\sqrt{C}} + \sqrt{1+\frac{(W+B)^2}{C}} \right|
-\ln \left|\frac{B}{\sqrt{C}} + \sqrt{1+\frac{B^2}{C}} \right|
\right] \nonumber \\
-\alpha &\left[ \sqrt{C+\left( W+B \right)^2} -\sqrt{C+B^2} \right]
\Biggr\}.
\label{eqn:same1}
\end{align}

An analytic expression for $b_0$ may be found in the interesting limit $\widetilde{\Delta} \to 0$, which is the situation that the two bands just touch at the $\Gamma$ point in the absence of coherence (i.e., for $b_0=0$.)  Taking this limit and dropping subleading terms in $\alpha b_0$, Eq. \ref{eqn:same1} simplifies to
$$
1 \approx \frac{Ug_0}{2}\left\{ \ln \frac{W}{b_0} -\alpha W \right\}
$$
which yields, for $U g_0 \ll 1$ and $\alpha W \sim 1$,
$$
b_0 \approx W \exp \left[-\frac{2}{Ug_0} \right],
$$
as stated in the main text.  We see in this case that the system supports spontaneous coherence even when there is no band crossing in it absence, provided the bands are sufficiently close in energy.

\subsection{Case 2: Crossing Bands of Opposite Chern number ($\mathcal{C}_{rel}=-1$)}

Recall from Eq. \ref{eqn:oppo} that the self-consistent equation for $B^{tb}$ for the case where the two crossing bands have opposite topology is
\begin{align}
B^{tb}({\bf k}_1)
&= \dfrac{U}{2V} \sum_{{\bf k}_2} (\cos^2 \theta_{{\bf k}_1}/2)(\cos^2 \theta_{{\bf k}_2}/2) \dfrac{B^{tb}({\bf k}_2)}{\sqrt{(\widetilde{\xi}({\bf k}_2)^2 + |B^{tb}({\bf k}_2)|^2}} \label{eqn:oppoBtb}
\end{align}
As a result, any self-consistent solution will be of the form $B^{tb}(\mathbf{k}) = b^2_0 \cos^2 (\theta_{\mathbf{k}}/2)$ where $b_0$ is a constant that does not depend on $\mathbf{k}$. Making the same replacement of $\theta_{\mathbf{k}}$ as in the previous subsection, we can modify Eq. \eqref{eqn:oppoBtb} into the form

\begin{align}
1 &=\left.\frac{U}{2} \int_{0}^{W} d \varepsilon g\left(\varepsilon\right) \frac{\cos ^4 \theta\left(\varepsilon\right)/2}{\left[\left(\varepsilon+\widetilde{\Delta}\right)^{2}+b_{0}\cos^4 \theta\left(\varepsilon\right)/2\right]^{1 / 2}}\right. \nonumber
\end{align}
As above, $\cos \theta \approx -1+\frac{k^2}{2m^2}$, so that
\begin{align}
\cos^2 \frac{\theta}{2} = \frac{1 + \cos \theta}{2} \cong \frac{k^2}{4m^2}
\implies \cos^4 \frac{\theta}{2} = \frac{k^4}{16m^4}. \nonumber
\end{align}
Again recalling $\varepsilon_{\bf k} \equiv E_{\bf k}-E_{{\bf k}=0}\sim k^2$ for ${\bf k}$ near the $\Gamma$ point,
we take $\cos^4 \theta\left(\varepsilon\right)/2 \to \dfrac{\alpha^2 \varepsilon^2}{4}$.

%

Using this low-energy form for the cosines,  noting that the integrand is strongly peaked around $\varepsilon = -\widetilde{\Delta}$ and taking $g\left(\varepsilon\right) \rightarrow g_{0}$
we obtain
\begin{align}
1 &=\frac{U}{8} g_0 \int_{0}^{W} d \varepsilon \frac{\alpha^2 \varepsilon^2}{\left[\left(\varepsilon+\widetilde{\Delta}\right)^{2}+{1 \over 4}\alpha^2 b^2_{0}\varepsilon^2\right]^{1 / 2}} \nonumber\\
&\approx \frac{U}{8} g_0 \widetilde{\Delta}^2 \int_{0}^{W} d \varepsilon \frac{\alpha^2}{\left[\left(\varepsilon+\widetilde{\Delta}\right)^{2}+{1 \over 4}\alpha^2 b^2_{0}\varepsilon^2\right]^{1 / 2}} \nonumber
\end{align}
The integral equation can be rewritten in the form
\begin{align}
1 & =\frac{\alpha^2 U g_{0}\widetilde{\Delta}^2}{8} \int_{0}^{W} d \varepsilon \frac{1}{\left[A(\varepsilon+B)^{2}+C\right]^{1 / 2}} \nonumber\\
& =\frac{\alpha^2 U g_{0}\widetilde{\Delta}^2}{8} \int_{B}^{W+B} d \varepsilon \frac{1}{\left[A\varepsilon^{2}+C \right]^{1 / 2}}, \nonumber
\end{align}
where $A = 1 + {1 \over 4}\alpha^2 b^2_0$, $B =  \frac{\widetilde{\Delta}}{1 + {1 \over 4}\alpha^2 b^2_0}$, and $C = {1 \over 4}\alpha^2 b^2_0 \frac{(\widetilde{\Delta})^2}{1 + {1 \over 4}\alpha^2 b^2_0}$.
Computing the integral generates the transcendental equation
\begin{align}
\frac{8}{\alpha^2 U g_0\widetilde{\Delta}^2} &= \ln \left(\frac{\sqrt{C+AW^2}+W\sqrt{A}}{\sqrt{C+AB^2}+B\sqrt{A}}\right),
\label{eqn:trans}
\end{align}
in which we have set $W+B \approx W$.

In contrast to the previous case, non-vanishing solutions for $b_0$ to Eq. \ref{eqn:trans} do not exist in the limit $\widetilde{\Delta} \to 0$.  A self-consistent solution can instead be found with
$\left|\widetilde{\Delta}\right| \gg  b_{0}^{2}$.
This is the statement that $b_0$ drops very rapidly as the bias changes, such that the Fermi surface that is present in the absence of coherence shrinks to a point.  With this assumption, $A\approx 1$, $B\approx \widetilde{\Delta}$, and $C \approx {1 \over 4}\alpha^2 b^2_0 \widetilde{\Delta}^2$. Notice that $B^2 \gg C$ and that we are interested in the case where the bands cross, so that $B<0$. Then
\begin{align}
\frac{8}{\alpha^2 U g_0\widetilde{\Delta}^2}
&= \ln \left(\frac{-4WB}{C}\right) \nonumber\\
\implies C &= -4W B \exp \left[-\frac{8}{\alpha^2 U g_0 \widetilde{\Delta}^2} \right] \nonumber\\
\implies b^2_0 &= -\dfrac{16W} {\alpha^2 \widetilde{\Delta}} \exp \left[-\frac{8}{\alpha^2 U g_0 \widetilde{\Delta}^2} \right]. \label{eqn:oppo_final}
\end{align}
This equation is equivalent to what is presented in the main text.
Note that Eq. \ref{eqn:oppo_final} yields a real value for $b_0$ when $\Delta < 0$.  As assumed above, we see that $b_0 \ll |\widetilde{\Delta}|$, and moreover $b_0=0$ for $\widetilde{\Delta}>0$.  In contrast to the case of $\mathcal{C}_{rel}=1$, the coherence gap vanishes precisely when the bands just touch, so that the topological transition and the $U(1)$ symmetry-breaking transition precisely coincide.



%
%
%
%

\bibliography{theSMbibliography}